\documentclass[%
reprint,
superscriptaddress,
 amsmath,amssymb,
 aps,
floatfix,
showkeys
]{revtex4-1}

\usepackage[usenames, dvipsnames]{color}
\usepackage{graphicx}
\usepackage{dcolumn}
\usepackage{bm}

\usepackage[section]{placeins}
\usepackage{url}
\usepackage{times}
\usepackage{textcomp}
\usepackage{multirow}
\usepackage{subcaption}
\usepackage{amsmath}

\begin{document}

\title{Low-frequency Raman signature of Ag-intercalated few-layer MoS$_2$}
\author{Natalya Sheremetyeva}
\affiliation{%
 Department of Physics, Applied Physics, and Astronomy, Rensselaer Polytechnic Institute, Troy, NY 12180, USA
}
\affiliation{%
 Thayer School of Engineering, Dartmouth College, Hanover, NH 03755, USA
}
\author{Drake Niedzielski}%
\affiliation{Department of Physics, Applied Physics, and Astronomy, Rensselaer Polytechnic Institute, Troy, NY 12180, USA
}
\author{Damien Tristant}%
\affiliation{Department of Physics, Applied Physics, and Astronomy, Rensselaer Polytechnic Institute, Troy, NY 12180, USA
}
\affiliation{Cain Department of Chemical Engineering, Louisiana State University, Baton Rouge, LA 70803, USA
}
\author{Liangbo Liang}
\affiliation{%
 Center for Nanophase Materials Sciences, Oak Ridge National Laboratory, Oak Ridge, TN 37831, USA
}
\author{Lauren E. Kerstetter}
\affiliation{%
Department of Materials Science and Engineering and Materials Research Institute, The Pennsylvania State University Park, State College, PA  16802, USA
}

\author{Suzanne E. Mohney}
\affiliation{%
Department of Materials Science and Engineering and Materials Research Institute, The Pennsylvania State University Park, State College, PA  16802, USA
}

\author{Vincent Meunier}
\affiliation{%
Department of Physics, Applied Physics, and Astronomy, Rensselaer Polytechnic Institute, Troy, NY 12180, USA
}%
\email{meuniv@rpi.edu}

\date{\today}

\begin{abstract}
Density functional theory based calculations and experimental analysis on a limited number of real samples are performed to study how the presence of silver intercalated in the van der Waals gap of few-layer MoS$_2$ affects the low-frequency Raman active modes of this material. Silver is found to predominantly affect the breathing-like and shear-like vibrational modes of MoS$_2$. These modes correspond to quasi-rigid movements of each individual layer with a restoring force (and, in turn, frequency) that is determined by modulations in the weak interlayer interactions. Noticeable red-shifts with increasing Ag concentration are found for all low-frequency modes. This finding indicates the potential for low-frequency vibrations as useful gauges for practical determination of silver concentration using low-frequency Raman spectroscopy. This work also describes a semi-classical linear chain model that allows to extrapolate results to a large number of layers. Further, first-principles calculations show how Raman spectroscopy can be used to characterize the quality of the two-dimensional interface between MoS$_2$ and a silver substrate.
\end{abstract}

\maketitle

\section{\label{sec:intro}Introduction}
Two-dimensional molybdenum disulfide is one of the most studied low-dimensional materials among the transition metal dichalcogenides (TMDCs) in the development of future nanoelectronic and optoelectronic applications thanks to its relatively large and tunable electronic bandgap~\cite{Radisavljevic2011}. The sizeable direct band gap and the large effective mass together with its thin 2D structure make few-layer MoS$_2$ attractive for low-power, scaled logic devices~\cite{Majumdar2014}. The thin body of the semiconductor makes it possible to circumvent short-channel effects in field-effect transistors. However, as in other nanoscale electronic devices, one key issue remains the control over the contacts with external circuitry. High contact resistance limits the performance of MoS$_2$-based devices, reducing the ON-state current and ON/OFF current ratio~\cite{Kuzum2011}. For MoS$_2$ to be usable in future nanoscale electronics, low-resistance Ohmic contacts need to be achieved and understood well. Reducing the contact resistance has been explored experimentally using many approaches~\cite{Rai2018,Schulman2018,Freedy2020}. One such method employs an annealing procedure performed to dope few-layer MoS$_2$ from a metallic contact such as silver~\cite{Abraham2017}. This approach has been used in the past for Ag contacts to bulk MoS$_2$~\cite{Souder1971,Souder1972} and has been revisited more recently for few-layer samples, leading to contact resistances as low as 0.2 kW $\mu$m~\cite{Abraham2017}. However, the detection of Ag diffusion and the exact characterization of Ag-intercalated MoS$_2$ remain challenging.

Raman spectroscopy has proven to be a particularly powerful tool for the characterization of two-dimensional materials (2DMs). Even small structural changes, \textit{e.g.}, the number of layers and their relative stacking order, can be detected with high precision by this non-destructive technique~\cite{Liang2017}. The Raman signature of pristine MoS$_2$ has been widely studied both experimentally and theoretically in various conditions such as the application of strain~\cite{Rice2013} or presence of internal defects~\cite{Parkin2016}. However, the Raman signature of Ag-intercalated MoS$_2$ has only recently been explored experimentally by Domask with the objective of elucidating the exact location of the Ag-impurities and their impact on the host's structure~\cite{Domask2017}. While the work of Domask provided valuable analyses of the HF Raman response of relatively thick MoS$_2$ samples (at least five layers), the low-frequency (LF) Raman signature and few-layer systems remains largely unexplored. 

Physical intuition suggests that, if Ag atoms reside in the van der Waals (vdW) gap, their effect should be most noticeable in the LF interlayer modes. The interlayer modes correspond to quasi-rigid movements of atomic layers as individual units and there are two types of such vibrations: The shearing modes (SM) are in-plane vibrations of layers parallel to each other and layer breathing modes (LBM) are out-of-plane vibrations~\cite{Liang2017,Zhang2016review}. The vibrational frequencies of these layer modes are mainly determined by restoring forces governed by the weak interlayer interactions. Thus, introducing external impurities in the space between the layers is expected to manifest itself as a shift in vibrational frequencies of these modes, since the interlayer interactions are expected to be strongly modified.

In this report, we study the vibrational properties of silver-intercalated few-layer MoS$_2$ using first-principles calculations based on density functional theory (DFT) with a focus on the LF interlayer vibrations. The HF modes are obtained as well as part of the complete analysis of the normal modes of the material and are reported for completeness in the Supplemental Information (SI) [link to SI]. All LF modes are found to red-shift significantly with increasing silver concentration. To compare predictions with experiments, the LF  Raman response in 2L and 3L MoS$_2$ was recorded experimentally and is reported here for the first time to provide experimental evidence for the predicted phenomenon. Moreover, the results obtained with DFT were used to calibrate a semi-classical model that allows for studying LF modes of systems with large numbers of layers. Finally, an alternative situation of placing a silver membrane on bilayer MoS$_2$ was explored computationally and revealed additional ultra-low frequency interface vibrational modes together with opposite shifts for the original SM and LBM. This study highlights the role of LF Raman spectroscopy for detailed characterization of the vibrational properties of Ag-intercalated MoS$_2$.

\section{\label{sec:methods}Methods}
\subsection{\label{sec:comptheo}Computation and theory}
First-principles calculations based on density functional theory were performed using the VASP package~\cite{Kohn1965, Kresse1996,Kresse1999}. VASP employs a plane-wave basis set and the projector augmented-wave (PAW) method to describe electron-ion interactions~\cite{Blochl1994}. Local-density approximation (LDA) was adopted for the exchange-correlation potential~\cite{Kohn1965}. For many layered 2DMs, LDA has been shown to provide a very good treatment of the lattice parameters and phonon frequencies compared to experiment~\cite{Liang2017}. This is partly because of fortuitous cancellations of errors, since LDA overestimates the covalent binding between layers while it does not capture the weak interlayer vdW forces. Few-layer MoS$_2$ (2-4 layers) was modeled by a periodic slab geometry. Vacuum regions of about 13~\AA~for 4L, 15~\AA~for 3L, and 17~\AA~for 2L in the out-of-plane direction were used to avoid spurious interactions with replicas. A plane-wave basis energy-cutoff of 600~eV and a dense Monkhorst–Pack~\cite{Monkhorst1976} 24$\times$24$\times$1 $k$-point grid sampling were found to be sufficient to accurately describe pristine MoS$_2$. The cell parameters and atoms were allowed to relax until the residual forces were below 1~meV/$\text{\AA}$ while the volume of the slab was kept fixed. 

Different silver concentrations, $C_{\mathrm{Ag}}$, were modeled using the supercell approach~\cite{Freysoldt2014}. In-plane supercells of pristine MoS$_2$ were constructed with extensions ranging from $2\times2$ ($C_{\mathrm{Ag}}=1.82\,$at\%) to $5\times5$ ($C_{\mathrm{Ag}} = 0.66\,$at\%) for all layer numbers and additionally for 2L, extensions $6\times6$ and $7\times7$ were considered. Such large supercells were computationally not feasible for 3L and 4L systems. A single Ag atom was placed at the octahedral (H) site in the middle of one vdW gap in each supercell. The H site was previously found to be energetically slightly more favorable compared to the tetrahedral (T) site for Ag in bulk MoS$_2$ in the $C_{\mathrm{Ag}}$-range considered here~\cite{Guzman2017}. The silver concentration is calculated as one Ag atom divided by the total number of atoms in the 2L supercell. Since only one vdW gap is Ag-intercalated in 3L and 4L, all systems have the same $C_{\mathrm{Ag}}$ per gap for the same in-plane supercell. 
Placing Ag atoms in only one vdW gap corresponds to an idealized situation. It allows us to first examine the trends of how Raman active modes are affected by Ag intercalation in the simplest possible case. We have then also checked the analogous mode renormalizations for 3L MoS$_2$ containing a moderate Ag-concentration in both gaps. The results of this simulation experiment are included in figures S4 and S5 and are referred to in the discussion in connection with comparisons to experiments.
After introducing the Ag-impurity, the atomic positions in each supercell system were re-optimized while keeping the cell geometry fixed until all residual forces were below 5 meV/\AA. Sufficiently fine $k$-space grids were chosen to obtain total energies converged within a couple of meVs for the smallest $3\times3$ supercell of 2L. 

Phonon frequencies and eigenvectors were obtained within the finite displacement method (FDM) in the harmonic approximation using the \texttt{Phonopy} package~\cite{Togo2015}. Additional appropriately sized supercells were used for phonon calculations in the smaller Ag-intercalated supercells to avoid unphysical interactions between periodic images. The convergence of the phonon frequencies was carefully tested with respect to the supercell sizes and was reached within few cm$^{-1}$. It should be noted here that DFT is formally a zero-temperature framework and thus no temperature effect is included in the present study.

\subsection{\label{sec:exp_methods}Experiments}
Electron beam evaporation was used to coat bulk MoS$_2$ crystals with 140~nm of Ag, on top of which 20~nm of SiO$_2$ was added to avoid dewetting of the Ag upon heating. Pieces were annealed at 400$^{\circ}$C or 500$^{\circ}$C for 24 hours in a tube furnace under flowing Ar gas. MoS$_2$ samples were mechanically exfoliated from below the silvered surface using thermal release tape and transferred to a SiO$_2$ (50 nm)/Si substrate, which had previously been patterned with Ti/Au tracking markers. Using optical contrast \cite{Li2012, Li2013}, a sampling of over twenty suspected few-layer flakes were identified under a microscope, and the individual flakes' locations recorded for subsequent Raman characterization. Additional details on sample preparation are included in the SI [link to SI].

Raman spectra were collected using a LabRAM HR Evolution Raman Spectrometer with excitation by the 532~nm (2.33~eV) line of a diode-pumped Nd:YAG solid-state laser using a grating with 1800~grooves/mm and a liquid nitrogen-cooled back-illuminated charge-coupled device detector. Laser-induced heating of the sample was avoided by operating at low power (1.44 mW). Each spectrum is an average of three 20-s acquisitions to improve the signal-to-noise ratio. The spectral resolution of this detector is $\sim$0.5 cm$^{-1}$. Great care was taken to ensure optimum laser focus and consistent data calibration. Spectra were collected from –100 to 550~cm$^{-1}$ to simultaneously measure both the Stokes and Anti-Stokes shifts. An ultra-low frequency filter enabled measurement within 10~cm$^{-1}$ of the Rayleigh line.  The spectra were then calibrated by making the Stokes and anti-Stokes SM peaks symmetric about the Rayleigh line. After fitting a 6th-order polynomial baseline to the data and subtracting the background intensity, we fit the Raman shifts using the Labspec6 peak fitting tool. A mixed Gaussian-Lorentzian shape was fit to each spectrum with a maximum of 25 iterations. Although peak fitting benefited from precise focusing of the laser, the weak LBM features still often required manual adjustment. The LF spectra were normalized to the SM.

\section{\label{sec:results}Results and Discussion}
Bulk MoS$_2$ belongs to space group P6$_3$/mmc (No. 194) with point group D$_{6h}$ (6/mmm). It has three Raman active modes: One LF interlayer shearing mode with $E_{2g}$ symmetry and two HF intralayer modes with symmetries $E_{2g}$ and $A_{1g}$. When scaled down to few layers, MoS$_2$ continues to exhibit a qualitatively similar Raman spectrum, but with additional LF interlayer breathing modes with $A_{1g}$ symmetry that are not present in the bulk~\cite{Liang2017}. The number of LF modes, their frequency $\omega$, and Raman activity depend on the symmetry of the few-layer system. Note that the symmetry of the few-layer systems is different from the bulk one (although related through group-subgroup relations) and depends on whether the number of layers $L$ is even or odd. Formally, the Raman active modes of few-layer MoS$_2$ have different irreducible representations compared to their bulk counterparts, because of the different symmetry groups. However, following a common practice in the literature and to avoid confusion, the irreducible representation labels of the bulk are used throughout this work, unless stated otherwise.

For example, $L=2$ has one SM and one LBM whose frequencies are known to follow opposite trends with increasing number of layers. Specifically, SM blue-shifts and saturates at the value that is found in bulk MoS$_2$ while LBM red-shifts and disappears for the bulk system~\cite{Zhao2013,Terrones2014,Liang2017}. These trends are well known and have been used as a practical gauge to determine the number of layers in experimental samples, see \textit{e.g.}, Ref.~\onlinecite{Liang2017} and references therein. 

\subsection{\label{subsec:vib_params}Vibrational modes in Ag-intercalated systems}
\begin{figure}[ht!]
\centering
\includegraphics[width=0.4\textwidth,keepaspectratio]
{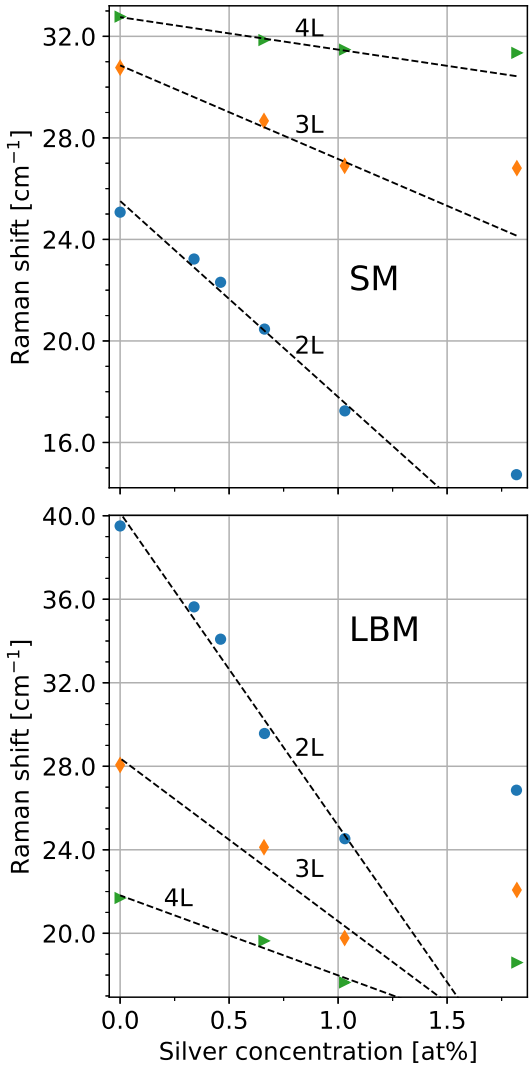}
\caption{\label{fig:app_chap4_LFmodes}
Calculated phonon frequency of the shearing (top) and breathing (bottom) modes as a function of silver concentration in two, three, and four layers of MoS$_2$ (see legend). Linear fits to the data points are indicated by dashed lines. In each system, a silver atom is located only in one (bottom) of the available interlayer gaps. Silver concentration is given in terms of the atomic concentration in 2L as the ratio between one Ag atom per gap over the total number of atoms in the 2L in-plane supercell. In this way, the in-plane Ag-concentration for a given supercell is the same for all layers. Linear fit results are summarized in table~\ref{tab:chap4_LFmodes}.
}
\end{figure}

Figure \ref{fig:app_chap4_LFmodes} shows the calculated $\omega \left( C_{\mathrm{Ag}} \right)$ dependence for the LF modes in 2L, 3L, and 4L MoS$_2$. Analogous results were found for the HF modes, albeit with a less pronounced effect of Ag intercalation, see Fig. S8 [link to SI]. In the rest of this work we focus on the LF interlayer modes. 

The known $\omega$ \textit{vs.} $L$ trends of pure MoS$_2$ are well reproduced here ($C_{\mathrm{Ag}}=0$ data points in figure~\ref{fig:app_chap4_LFmodes}), thus providing the baseline for studying the effect of Ag intercalation. For all numbers of layers considered here, the frequencies of the LF modes behave approximately linearly up to about $C_{\mathrm{Ag}}=1\,$at\% and exhibit a red-shifting trend. At 2~at\%, the LF-modes' frequencies reverse their trends and blue-shift compared to their values at a lower silver concentration. The origins of this trend-reversal will be discussed in more detail below, but it should be noted that experimental samples are expected to have Ag-concentrations well below 1 at\%~\cite{Domask2017}. In the red-shift region, linear functions can be fit to the $\omega(C_{\mathrm{Ag}})$ curves (dashed lines in figure~\ref{fig:app_chap4_LFmodes}) to quantitatively assess the effect of Ag for different numbers of MoS$_2$ layers. The resulting $\partial \omega/ \partial C_{\mathrm{Ag}}$ slopes are summarized in table~\ref{tab:chap4_LFmodes}. Overall, the LBM is more sensitive to $C_{\mathrm{Ag}}$ than its SM counterpart in a given system with the absolute fitted slopes for LBM being about a factor of two larger than the corresponding ones of SM. Moreover, both LBM and SM become less sensitive to $C_{\mathrm{Ag}}$ with increasing number of layers as their corresponding $\partial \omega/ \partial C_{\mathrm{Ag}}$ slopes decrease for each layer added to the structure. The smaller frequency shifts in thicker samples imply that it would become more difficult to detect the presence of Ag atoms in these samples compared to thinner ones. Other potential limitations of the direct comparison between predictions and experiments will be discussed in more detail in Sec. \ref{sec:limits}.

\begin{table}[]
\centering
\begin{ruledtabular}
\begin{tabular}{cccc}
Mode & 2L              & 3L             & 4L             \\
\hline
SM   & -7.7 $\pm$ 0.6  & -3.7 $\pm$ 0.4 & -1.3 $\pm$ 0.1 \\
LBM   & -15.0 $\pm$ 1.1 & -7.8 $\pm$ 1.5 & -3.8 $\pm$ 0.6 \\
\end{tabular}
\end{ruledtabular}
\caption{\label{tab:chap4_LFmodes}Linear fit results for the $\partial \omega/ \partial C_{\mathrm{Ag}}$ slopes in units of cm$^{-1}$/at\%. The slope uncertainty is given by the standard error of the linear regression.
}
\end{table}
The frequencies of the SM and LBM reported in Fig.~\ref{fig:app_chap4_LFmodes} correspond to the ones that are typically detected experimentally in pure MoS$_2$ and are now tracked as a function of $C_{\mathrm{Ag}}$ in Ag-intercalated systems via a phonon-unfolding procedure described in more detail in the SI [link to SI]. However, generally there are $L-1$ doubly degenerate shearing and $L-1$ breathing modes in a system with $L$ layers~\cite{Liang2017}. Depending on the system's symmetry and the symmetry of the atomic displacements in a given mode, these additional modes can be Raman inactive, or they can be Raman active, but undetectable in a typical experimental back-scattering set-up due to the form of the corresponding Raman tensor. The introduction of Ag in the system can change the overall symmetry and, in turn, the Raman activity of the normal modes. More detailed symmetry considerations for each system are given in the SI [link to SI].

\begin{figure}[ht!]
\centering
\includegraphics[width=0.35\textwidth,keepaspectratio]
{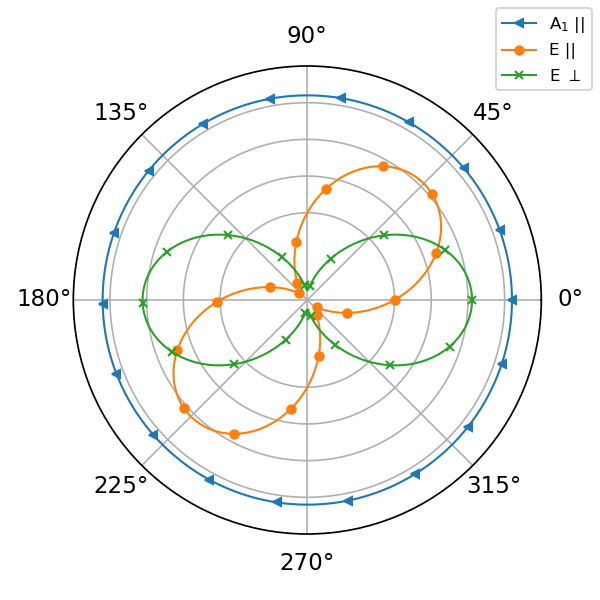}
\caption{\label{fig:pol_raman_profiles}
Theoretical polarization-angle resolved Raman intensity profiles for modes $A_1$ and $E$ based on the Raman tensors provided by equation~\ref{eq:RT} for two typical laser configurations: Parallel laser configuration (denoted $||$) and cross laser configuration (denoted $\perp$). The Raman signal of the $A_1$ mode vanishes for cross laser configuration. For illustration, the non-zero elements of the Raman tensors were assumed to be real valued and were set arbitrarily to $a=0.8$ and $c=0.65$. The mathematical details of how these profiles were computed can be found in SI [link to SI]. Note that due to the resulting formulas simplifying to $I(E, ||) \propto |c|^2 \sin^2(2\theta)$ and $I(E, \perp) \propto |c|^2 \cos^2(2\theta)$, the two polarization profiles have a phase angle of 45$^{\circ}$ between them.
}
\end{figure}

The phonon frequencies of the LF interlayer modes in 4L and their irreducible representation (IR) labels are tabulated in table~\ref{tab:4L_sym_freqs} for pure and Ag-intercalated MoS$_2$ with $C_{\mathrm{Ag}}$ of $\sim$ 1 at\% (4$\times$4 supercell). The IR labels provide information about Raman activity. An analogous overview for 2L and 3L, associated Raman tensors and visualizations of underlying atomic displacements are provided in figures S3 through S7 [link to SI]. The modes $E_g$(2) and $E$(3) correspond to the SMs and $A_{1g}$(1) and $A_{1}$(1) are the LBMs in Fig.~\ref{fig:app_chap4_LFmodes}. First, all LF modes red-shift with $C_{\mathrm{Ag}}$, but the magnitudes of the shifts for different SMs and LBMs are not the same as the ones observed and provided in table~\ref{tab:chap4_LFmodes}. This is expected because the details of the atomic displacements in a given mode determine how this mode's phonon frequency changes with varying external or internal conditions that cause a global change in the interatomic force-constants. The interplay between atomic-displacement patterns and global changes of the interatomic force-constants can sometimes even lead to quite counter-intuitive shifts in frequency~\cite{Sheremetyeva2018}. Second, the modes $E_u$(1) and $A_{2u}$(1) are Raman inactive in pure 4L MoS$_2$ but are activated due to a reduction in symmetry induced by Ag-intercalation and now correspond to modes $E$(2) and $A_{1}$(2), respectively. The general shape of the Raman tensors of the latter modes is given by 
\begin{align}
 \widetilde{R}\left(A_{\mathrm{1}}\right)&=\left(\begin{array}{lll}
a & 0 & 0 \\
0 & a & 0 \\
0 & 0 & b
\end{array}\right),~ 
\widetilde{R}\left(E(x)\right)&=\left(\begin{array}{lll}
0 & c & d \\
c & 0 & 0 \\
d & 0 & 0
\end{array}\right), 
\label{eq:RT}
\end{align}
based on group theoretical considerations~\cite{Kroumova2003}. The Raman tensors indicate that these newly activated modes are in principle detectable in a typical Raman experiment with a back-scattering laser-polarization set-up. Lastly, note the highlighted phonon frequencies in table~\ref{tab:4L_sym_freqs}, which include the symmetry-activated modes in the Ag-intercalated system: The difference between the frequencies of the $A_{1g}$(1) and $E$(2) modes is below 1~cm$^{-1}$, and just 0.3 cm$^{-1}$ between $E_g$(2) and $A_{1}$(2). Such small differences can be challenging to detect experimentally, particularly since the latter is below the typical experimental resolution of the associated Raman-peak position of about 0.5~cm$^{-1}$~\cite{Domask2017}. Thus, in practice the recorded Raman spectra could appear very similar for pure and Ag-intercalated 4L MoS$_2$, especially if the other, further red-shifted modes cannot be clearly observed due to their increased proximity to the strong Rayleigh line. However, the pairs of the highlighted phonon modes belong to different irreducible representations, \textit{i.e.}, the symmetries of the underlying atomic displacements are different between $A$ (LBM) and $E$ (SM) type modes. In this case, polarization-resolved Raman spectroscopy could help determine the mode-symmetry associated with a given Raman peak and thus distinguish between pure and Ag-intercalated MoS$_2$. Polarized Raman spectroscopy has been successfully used for mode-symmetry assignment in pure MoS$_2$ and other 2D materials~\cite{Chen2015,Cao2017}, but has not been explored in Ag-intercalated MoS$_2$ to date. Figure~\ref{fig:pol_raman_profiles} shows the polarization-angle resolved Raman profiles that can be expected for the modes in question based on the general shape of their associated Raman tensors in equation~\ref{eq:RT}. We note that the profiles shown in Fig.~\ref{fig:pol_raman_profiles} were calculated based on the assumption of the non-zero elements of the Raman tensors in equation~\ref{eq:RT} to be real valued and have the values of $a=0.8$ and $c=0.65$ for illustration purposes. In experiments, resonance effects might become important and might affect the exact values of the non-zero elements of the Raman tensors of the $A_1$ and $E$ modes (elements can become complex valued at resonance conditions). However, the general shape of the polarization profiles is not expected to change with the laser energy. Thus, the very different shapes of the profiles for the $A_1$ and $E$ modes in principle allows for their differentiation at different laser energies making the profiles in Fig.~\ref{fig:pol_raman_profiles} a potentially useful reference for future experimental work.

\begin{table}
\centering
\begin{ruledtabular}
\begin{tabular}{cccc}
\multicolumn{2}{c}{Pure} & \multicolumn{2}{c}{$C_{\mathrm{Ag}}=1\,$at\%}  \\  \cline{1-2} \cline{3-4}
Mode & $\omega$ [cm$^{-1}$] & Mode & $\omega$ [cm$^{-1}$] \\
$E_g$(1) & 13.4 & $E$(1) & 11.6 \\
$A_{1g}$(1) & \textbf{21.7} & $A_{1}$(1) & 17.6 \\
$E_u$(1) & 25.0 & $E$(2) & \textbf{21.0} \\
$E_g$(2) & \textbf{32.8} & $E$(3) & 31.5 \\
$A_{2u}$(1) & 39.9 & $A_{1}$(2) & \textbf{32.5} \\
\end{tabular}
\caption{Calculated phonon frequencies of the LF modes in pure and Ag-intercalated 4L-MoS$_2$ with $C_{\mathrm{Ag}}=1\,$at\% together with the associated irreducible-representation labels based on group theory. The latter are assigned via symmetry-analysis of the corresponding phonon eigenvectors, which is provided within the \texttt{Phonopy} package~\cite{Togo2015}. Here, the irreducible-representation labels of the actual corresponding point-groups were used instead of those of bulk MoS$_2$.}
\label{tab:4L_sym_freqs}
\end{ruledtabular}
\end{table}

The results presented thus far provide potentially useful gauges for the practical determination of $C_{\mathrm{Ag}}$ up to about 1~at\% using LF Raman measurements. However, in addition to the practical usefulness of the findings, their physical origins are interesting as well. Generally speaking, the addition of an impurity changes the inter-atomic force-constants as well as the total mass of the system and thus leads to changes in vibrational frequencies. For a simple harmonic oscillator, the frequency $\omega$ is given by $\omega=\sqrt{K/m}$ with the force constant $K$ and mass $m$. Thus, depending on the rate and direction of the change in $K$ relative to the increase in $m$, a certain normal mode can red- or blue-shift by a different amount. Next, we analyze possible pathways to the changes in the interlayer force-constants that cause the observed $\omega \left( C_{\mathrm{Ag}} \right)$ by examining the structural and electronic changes induced by silver intercalation into MoS$_2$.

\subsection{\label{subsec:struct_params}Structural Parameters}
\begin{figure}
\centering
\includegraphics[width=0.45\textwidth,keepaspectratio]
{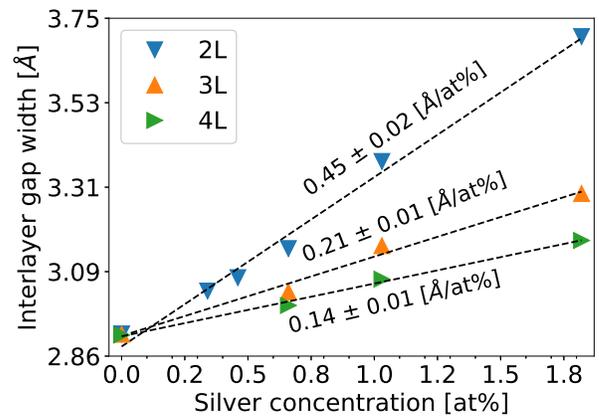}
\caption{\label{fig:app_chap4_avgap_allL}Calculated average interlayer-gap width, averaged over the in-plane supercell area and the number of gaps in the system, as a function of Ag concentration for different numbers of layers. Linear fit functions are indicated by dashed lines and the corresponding slopes are provided along the curves. The indicated uncertainty is the standard error of the linear fit.}
\end{figure}

Figure \ref{fig:app_chap4_avgap_allL} shows the average relaxed interlayer distance ($d_{\mathrm{gap}}$) as a function of silver concentration in all the few-layer systems. The average was taken over the in-plane area of the supercell in each vdW gap and over the number of gaps. 
The calculated interlayer distance of pure MoS$_2$ is about $2.92\,$\AA~ for all layer numbers in good agreement with the experimental value of about 3~$\text{\AA}$~\cite{Dong2018} and previous theoretical results~\cite{Xiao2014}. In all systems, $d_{\mathrm{gap}}$ increases in good approximation linearly as a function of $C_{\mathrm{Ag}}$ in the whole range of considered Ag-concentrations, but the increase rates are different depending on the number of layers. Similarly to the analysis provided in Fig.~\ref{fig:app_chap4_LFmodes}, the data here was fit to linear functions with the resulting $\partial d_{\mathrm{gap}}/\partial C_{\mathrm{Ag}}$ slopes indicated along the curves presented in Fig.~\ref{fig:app_chap4_avgap_allL}. Clearly, the rate of the interlayer-gap widening decreases with increasing number of layers. Moreover, the slope of 3L (4L) is essentially the slope of 2L divided by a factor of 2 (3), with the factor corresponding to the number of vdW gaps in the 3L (4L) material. This hints at the effect of Ag being confined to the interlayer gap in which the impurity has been placed resulting in $\partial d_{\mathrm{gap}}/\partial C_{\mathrm{Ag}}(L)$ being a simple average of the value for 2L over the number of gaps in the system.

Table~\ref{tab:app_chap4_slopesallL} provides linear-fit slopes analogous to those shown in Fig.~\ref{fig:app_chap4_avgap_allL}, but with the interlayer distance evaluated for each vdW gap individually (averaged over the in-plane extension of the corresponding unit cell). The corresponding gap-width visualizations are provided in figure S9 [link to SI]. For all layer numbers, the $\partial d_{\mathrm{gap}}/\partial C_{\mathrm{Ag}}$ slope for the in-plane-average interlayer-distance in the Ag-containing gap is that of 2L. Moreover, the gaps without silver in 3L and 4L remain practically unchanged with Ag-concentration, which confirms the locality of the structural changes. 
\begin{table}[]
\centering
\begin{ruledtabular}
\begin{tabular}{ccccc}
 & Bottom gap (with Ag) &	Middle gap (no Ag) &	Top gap (no Ag)\\
\hline
2L &	0.45 $\pm$ 0.02 &	- &	-\\
3L	&0.43 $\pm$ 0.03&	-0.01 $\pm$ 0.01 &	-\\
4L	&0.42 $\pm$ 0.02 &	-0.006 $\pm$ 0.002 &	-0.001 $\pm$ 0.003\\
\end{tabular}
\end{ruledtabular}
\caption{\label{tab:app_chap4_slopesallL}Results of linear fits of the $\partial d_{\mathrm{gap}}/ \partial C_{\mathrm{Ag}}$ slopes in units of \AA/at\% for each individual vdW gap in two, three, and four layers MoS$_2$. The uncertainty on the values of the slopes corresponds to the standard error of the linear fit.}
\end{table}

Having understood the nature of the scaling of the overall average $\partial d_{\mathrm{gap}}/\partial C_{\mathrm{Ag}}$ slopes as a function of the number of layers in Fig.~\ref{fig:app_chap4_avgap_allL}, the slopes can now be used to gain a qualitative understanding of the observed frequency red-shift trends shown in Fig.~\ref{fig:app_chap4_LFmodes}. The Ag-impurity pushes the MoS$_2$ layers apart resulting in an increased interlayer separation. This in turn corresponds to an effective softening of the interlayer force-constants. Together with the increased overall mass of the system, the presence of Ag manifests itself in the observed red-shift of the LF modes. The LBM is more sensitive to $C_{\mathrm{Ag}}$ than the SM, because the out-of-plane vibrations in LBM probe the interlayer force-constant in that direction. Similarly, since the Ag impurity has a local effect on the interlayer distance, the force-constant softening occurs only locally in the gap in which Ag is located. Thus, with increasing number of layers, the local force-constant softening effect is averaged out over the number of gaps that remained unaffected. As a result, the vibrational frequencies red-shift at a slower rate as $C_{\mathrm{Ag}}$ increases. 

The above discussion of the $\partial d_{\mathrm{gap}}/\partial C_{\mathrm{Ag}}$ slopes aids in rationalizing the red-shifting $\omega \left( C_{\mathrm{Ag}} \right)$ trends shown in Fig.~\ref{fig:app_chap4_LFmodes} up to about 1 at\%. The sudden blue-shift trends at $C_{\mathrm{Ag}} \approx 2\,$at\% is, however, counter-intuitive in view of the continuous vdW-gap widening in the whole Ag-concentration range shown in Fig.~\ref{fig:app_chap4_avgap_allL}. We conclude there is an additional mechanism at work that counteracts the effective force-constant softening associated with the vdW gap widening and leads to the observed blue-shift trend. Next, we examine the effect of charge-density redistribution upon silver intercalation as a possible reason accounting for this effect.
\subsection{\label{subsec:bader}Charge density analysis}
\begin{figure}
\centering
\includegraphics[width=0.45\textwidth,keepaspectratio]
{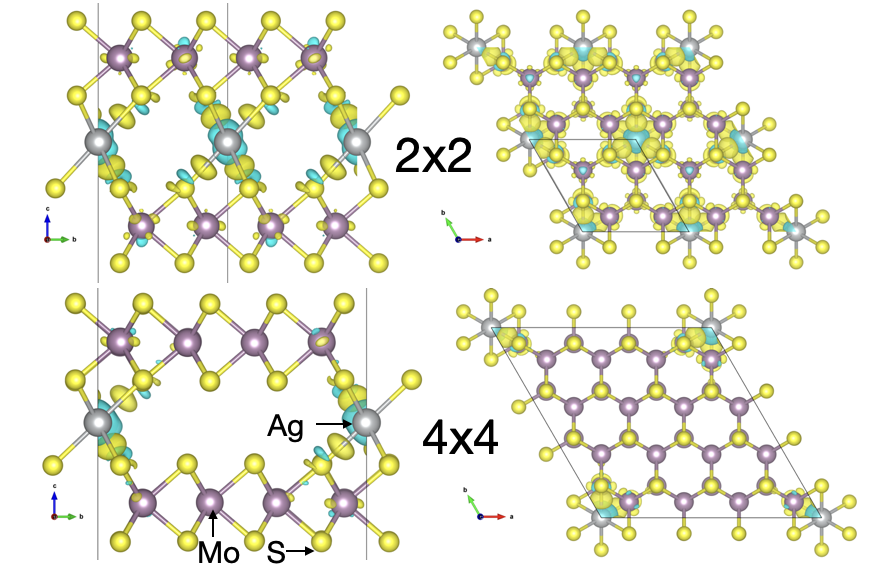}
\caption{\label{fig:charge_diff}Charge density difference plots for the $2\times2$ ($C_{\mathrm{Ag}}=1.8\,$at\%) and $4\times4$ ($C_{\mathrm{Ag}}=0.7\,$at\%) supercells of Ag-containing 2L MoS$_2$. The charge density difference is taken between the silver-intercalated, pristine MoS$_2$, and isolated Ag atoms placed at corresponding positions. The unit cell (marked by lines for each system) of the $2\times2$ supercell is periodically repeated once in plane for a better comparison to the $4\times4$ supercell on the same scale. The charge density difference indicates the redistribution of charge between the intercalated Ag in MoS$_2$ compared to when each individual system is isolated. The charge-density plots were made using the VESTA visualization tool~\cite{Momma2011}. For both supercells, the isosurface level was set to 0.0025 e/Bohr$^3$. The color scheme indicates positive (yellow) and negative (turquoise) values of the charge-density difference.
}
\end{figure}
The preceding discussion provided useful handles for the determination of $C_{\mathrm{Ag}}$ in few-layer MoS$_2$ using non-destructive Raman spectroscopy. The calculated $\omega \left( C_{\mathrm{Ag}} \right)$ behavior is linear in the range of Ag-concentrations that can be realistically expected in experiments. Nevertheless, it is interesting to study the physical origins of the $\omega \left( C_{\mathrm{Ag}} \right)$ trend-reversal when $C_{\mathrm{Ag}}$ increases well beyond 1 at\%. Intuitively, we expect to see the emergence of a mechanism opposing the continued effective force-constant softening brought on by continued vdW gap widening due to the presence of Ag impurities. One such mechanism could be the formation and further strengthening of covalent interlayer bonds mediated by Ag. To investigate this possibility, the charge-density differences in 2L MoS$_2$ intercalated with 1.8 at\% (top panel) and 0.7 at\% (bottom panel) Ag are examined in Fig.~\ref{fig:charge_diff}.

The charge-density difference is calculated between the silver-intercalated, pristine MoS$_2$, and isolated Ag atoms placed at appropriate positions. The unit cell (marked by lines for each system) of the $2\times2$ supercell is periodically repeated once in plane for comparison to the $4\times4$ supercell on the same scale. The charge-density difference indicates the redistribution of charge between the Ag impurity in MoS$_2$ compared to when each individual system is isolated. 

The charge density redistribution around the intercalated Ag is very similar in both systems. The charge density is depleted around the silver atom, but accumulated half-way along the bonds with neighboring S atoms. This finding is confirmed quantitatively by a Bader charge analysis~\cite{Henkelman2006}, where in both systems a similar charge of 10.66$e$ and 10.63$e$ is found in the Bader volume around Ag in the $2\times2$ and $4\times4$ supercells, respectively. Moving away from the Ag impurity, in the $4\times4$ supercell, the charge density does not change appreciably between pristine and Ag-intercalated MoS$_2$ systems. This cannot be seen in the $2\times2$ supercell, because of the large in-plane density of Ag impurities. In this situation, silver forms an extra layer in the vdW gap of MoS$_2$. In turn, the formation of the extra Ag-layer accounts for the effective strengthening of the interlayer interactions in MoS$_2$ that manifests itself in a hardening of the interlayer mode frequencies compared to a smaller silver concentration observed in the previous section.

\subsection{\label{subsec:linchain}Semi-classical Linear-Chain Model}

Up to this point, we described silver-intercalated few-layer MoS$_2$ systems by means of DFT calculations. We found that the Raman signature changes most dramatically with Ag-concentration in the bilayer system due to substantial structural and electronic changes induced by the presence of silver. Additionally, the Ag-induced changes are mostly local and affect primarily the vdW gap in which the silver impurities reside, explaining why the effects subside in three and four-layer MoS$_2$. The calculated $\omega \left( C_{\mathrm{Ag}} \right)$ trends also provide useful gauges for $C_{\mathrm{Ag}}$-determination in few-layer MoS$_2$ using Raman spectroscopy. It is desirable to extend the present analysis beyond few-layer MoS$_2$ systems to thicker samples with 5, 10 and more layers, as they are commonly fabricated~\cite{Abraham2017}. Unfortunately, such an extension is impossible in practice within the framework of DFT due to the large supercell sizes involved. 

For this reason, in this section we turn to a semi-classical approach to study the LF modes in many-layer MoS$_2$ intercalated with silver. As mentioned before, in the LF modes, the individual layers move rigidly as single units such that the vibrations are mainly governed by a few parameters describing the interlayer interactions. The atomistic details within the layers can thus be neglected in good approximation. The nearly classical nature of the LF modes' vibrations makes them suitable to be well described by means of the linear chain model (LCM)~\cite{Zhang2016review,Liang2017}. In the LCM, each layer is approximated as a rigid ball with mass density $\mu$ (units of mass per unit area) connected by springs with spring constants $\kappa$ (units of force per length per unit area) to its nearest neighbors. The normal modes are obtained from the diagonalization of the corresponding dynamical matrix of a system of $L$ coupled layers represented by massive particles. Within this picture, both the inter-layer shearing and breathing modes are topologically identical and their differences are included by considering different spring constants for each type of modes.  

In pure MoS$_2$, the spring constants between all layers are identical (for a given LF mode type -- SM or LBM). In turn, silver intercalation can be simulated by including a modified spring constant $\kappa'=\kappa+\mathrm{d}\kappa$ in one or more vdW gaps of the original system. This results in a modified dynamical matrix with a perturbation $\mathrm{d}\kappa$, which has to be numerically diagonalized to obtain the eigenmodes of the new Ag-perturbed system. 

Here, the interlayer force-constants $\kappa$ were first accurately calibrated to reproduce the DFT results of the SM and LBM frequencies in pristine few-layer MoS$_2$. The fitted force-constants are given by $\kappa^{\parallel}=3.34 \times 10^{19}$ N/m$^3$ (in-plane force-constant for the SM) and $\kappa^{\perp}=8.45 \times 10^{19}$ N/m$^3$ (out-of-plane force-constant for the LBM), in good agreement with previously reported values based on experimental LF modes frequencies \cite{Liang2017}. Next, the force-constant perturbations $\mathrm{d}\kappa$ were analogously fitted to our DFT results for different Ag concentrations. For this step, we assumed that only one spring-constant has to be modified to $\kappa'$, namely the one in the vdW-gap containing silver. This assumption is based on the earlier finding of Ag affects MoS$_2$ mostly locally. The resulting $d\kappa$ vs. $C_{\mathrm{Ag}}$ values were found to be practically identical for all number of layers. Having fully calibrated the LCM, we can now turn to systems made up of large numbers of layers.
\begin{figure}
        \centering
        \includegraphics[width=0.45\textwidth,keepaspectratio]{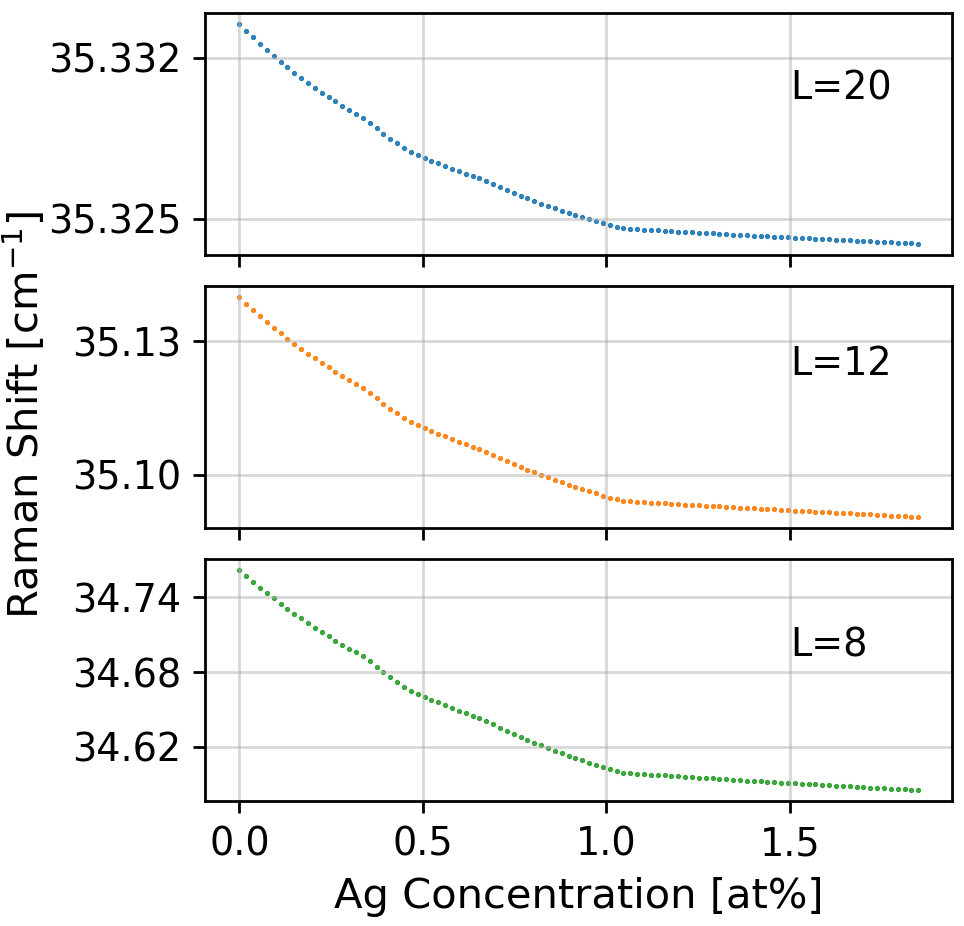}
\caption{Calculated frequency of the highest-frequency SM in 8, 12, and 20 -layer MoS$_2$ as a function of silver concentration obtained from LCM. Only one vdW gap contains silver impurities in each system.}
        \label{fig:LCM_shear}
\end{figure}
Figure~\ref{fig:LCM_shear} shows the frequency of the highest-frequency SM as a function of silver concentration for systems containing 8, 12, and 20 layers of MoS$_2$, where silver was placed in only one of the vdW gaps. Analogous results were found for the LBM, see Fig.~S11 [link to SI]. According to the predictions from the bond-polarizability model for pure MoS$_2$ \cite{Liang2017a}, the highest-frequency modes are predicted to have the largest Raman intensity, see Fig.~S10 for the calculated spectra [link to SI].

The $\omega \left( C_{\mathrm{Ag}} \right)$ trends are qualitatively the same for all numbers of layers since the trends are essentially inherited from DFT calculations in few-layer systems due to the fitting of the force-constants. However, note the range in which the SM-frequency red-shift takes place for each $L$: the $\Delta \omega$ red-shift from zero to maximal $C_{\mathrm{Ag}}$ is within a tenth of a wavenumber for $L=8$ and decreases significantly with increasing number of layers. This clearly demonstrates the eventual saturation of the $\partial \omega/ \partial C_{\mathrm{Ag}}$ shifts at zero for large $L$. Experimentally, such minute frequency changes would be practically impossible to resolve, resulting in the apparent absence of any change of the Raman signature upon Ag intercalation.  
\begin{figure}
        \centering
        \includegraphics[width=0.45\textwidth,keepaspectratio]{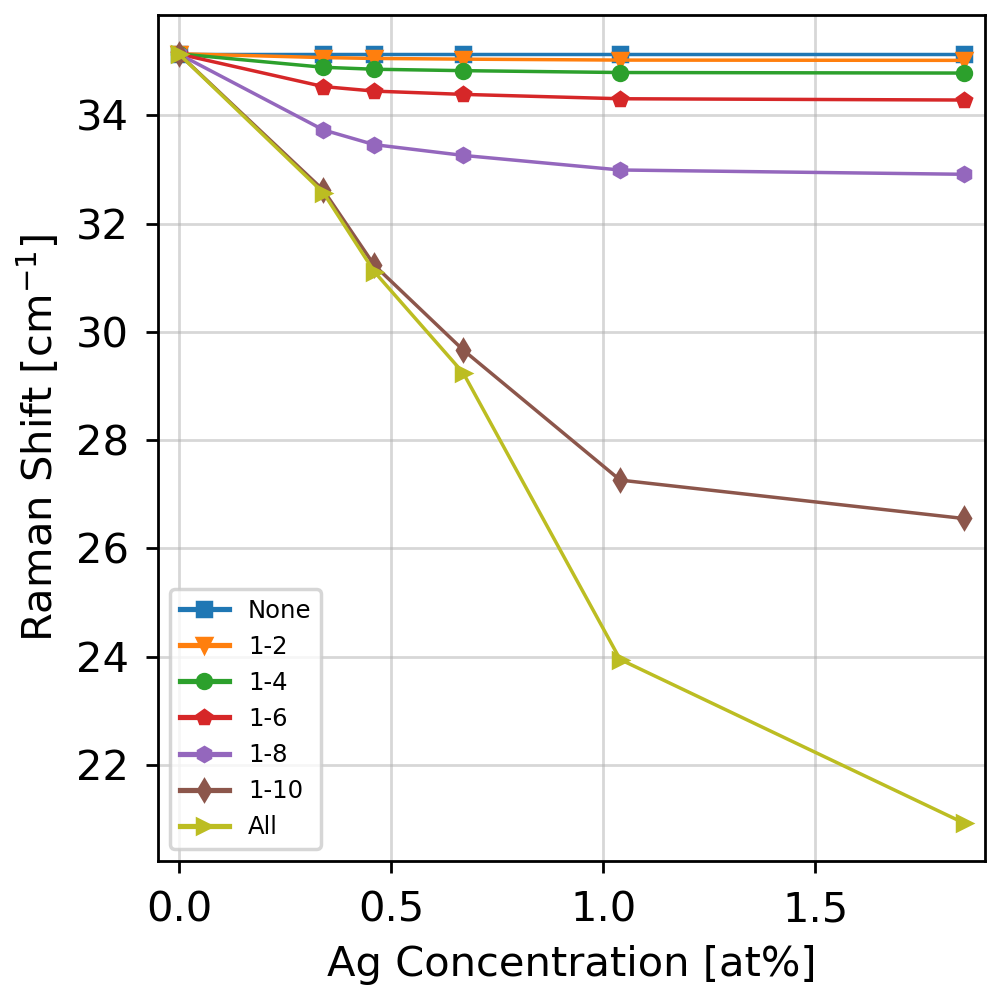}
\caption{This figure is similar to that shown in Fig.~\ref{fig:LCM_shear}, but here it is plotted for 12 layers only and after placing silver in a varying number of vdW gaps within the LCM. This corresponds to setting multiple spring-constants to perturbed values $\kappa'$ (see legend). The curve labeled "None" corresponds to the baseline of pure 12-layer MoS$_2$. It is followed by the case where only one spring-constant is modified (repeated from Fig. \ref{fig:LCM_shear}) labeled as "1-2", meaning that the vdW gap between layers 1 and 2 contains silver, and so on increasing the number of gaps between layers that contain the intercalated Ag atoms.}
        \label{fig:Multi_kappa_shear}
\end{figure}

Alternatively, one can fix the number of layers, but vary the number of gaps that accommodate silver atoms. This is an important setting to explore since experimentally the out-of-plane penetration depth of silver through MoS$_2$ is not easily controlled and determined~\cite{Domask2017}. Figure~\ref{fig:Multi_kappa_shear} shows the case of a fixed number of layers, $L=12$, and varying number of modified spring-constants $\kappa'$. The starting point is the SM frequency in pristine 12-layer MoS$_2$ at about 35~cm$^{-1}$. It corresponds to a system without any modified force-constants. For all the following curves, the number of Ag-containing gaps was gradually increased until all vdW gaps contain silver impurities (12-layer MoS$_2$ has 11 vdW gaps in total). For all cases, the SM red-shifts with $C_{\mathrm{Ag}}$, as before, but the red-shift rate increases with the number of gaps described by the modified spring-constant $\kappa'$. When all but one spring-constants are perturbed, the $\partial \omega/ \partial C_{\mathrm{Ag}}$ slope (fitted for $C_{\mathrm{Ag}}$ up to 1 at\%) is given by -7.66~cm$^{-1}/$at\%, thus practically recovering the value from the 2L case earlier, see table~\ref{tab:app_chap4_slopesallL}. When all force-constants are modified, the red-shift from zero to $C_{\mathrm{Ag}}=1\,$at\% is maximal with -10.74~cm$^{-1}/$at\%. 

However, note that for small Ag-concentrations the red-shift magnitudes are very similar between systems with different numbers of modified force-constants. This implies that in order for the calculated changes to be useful for comparison to experiment, the experimental Ag-penetration depth would have to be determined by some other means. Otherwise it would be difficult to extract both the number of gaps containing Ag and the Ag-concentration within the gaps, by means of one Raman measurement.
\subsection{\label{subsec:substrate}Ag-membrane on bilayer MoS$_2$}
Up to this point, we have discussed the effect of silver intercalation into the interlayer gaps of MoS$_2$. Another experimentally possible setting where Ag interacts with MoS$_2$ is the deposition of a thick silver membrane on MoS$_2$. This situation can precede the interlayer intercalation. For example, one experimental pathway to achieve Ag-intercalation is to first deposit a silver membrane, then anneal the complete system and allow Ag to diffuse through the MoS$_2$ layers, and finally exfoliate few-layer Ag-intercalated MoS$_2$. To analyse this effect, we turn back to a DFT-based analysis and discuss the vibrational properties at a silver/MoS$_2$ interface.
\begin{figure}[h!]
\centering
\includegraphics[width=0.47\textwidth,keepaspectratio]{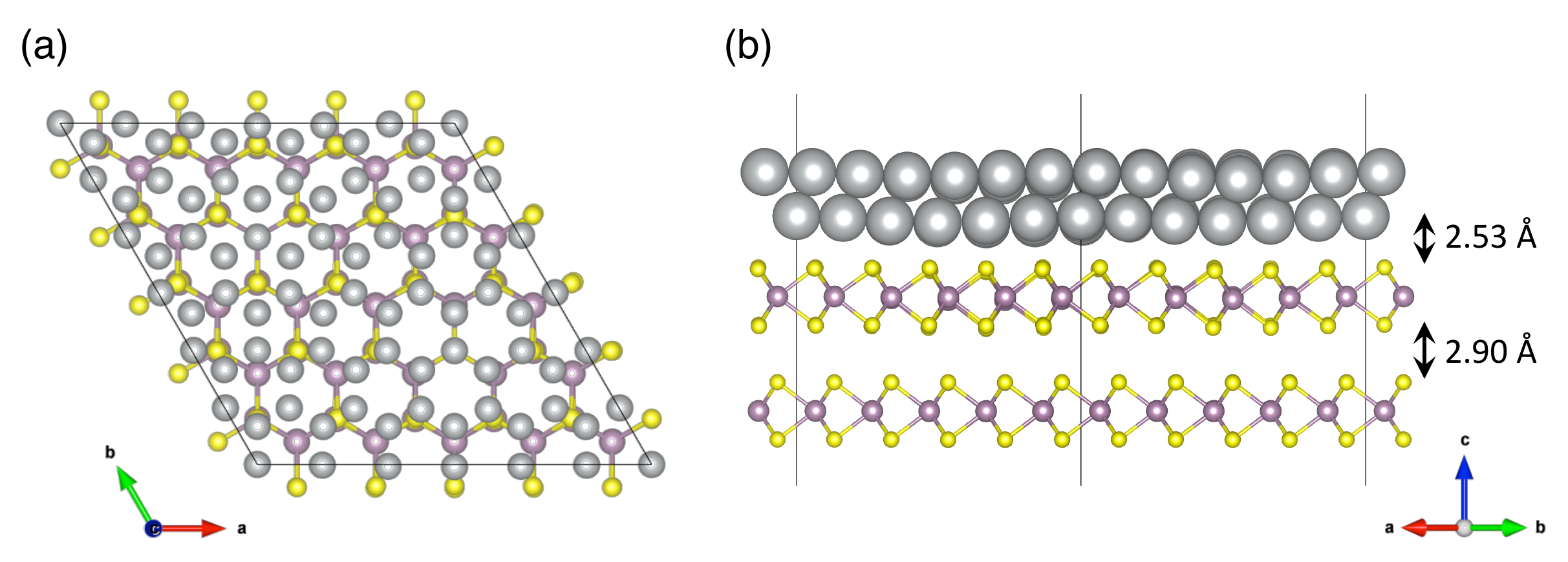}
\caption{(a) Top and (b) side views of Ag(111) adsorbed on bilayer MoS$_2$. The unit cell of the composite nanostructure is comprised of a $(5\times 5)$ in-plane supercell of pristine MoS$_2$ and a $(6\times 6)$ repetition of Ag(111). Different supercell sizes were considered to create the best possible match between cells. The configuration shown here was found to be the most energetically favorable, see Fig.~S13 [link to SI].}
\label{Fig:structure}
\end{figure}
A two-atom-thick layer of Ag(111) was adsorbed on 2L MoS$_2$, as shown in Fig.~\ref{Fig:structure}. Different lattice extensions of the Ag(111) and MoS$_2$ lattices were analyzed to construct a composite nanostructure with the best possible lattice match, see SI for technical details [link to SI]. As a result, a $(5\times 5)$ supercell of MoS$_2$ together with a $(6\times 6)$ supercell of Ag(111) was found to be the most energetically favorable configuration and was used for subsequent investigation of phonon modes in this system. 
\begin{figure}[h!]
\centering
\includegraphics[width=0.47\textwidth,keepaspectratio]{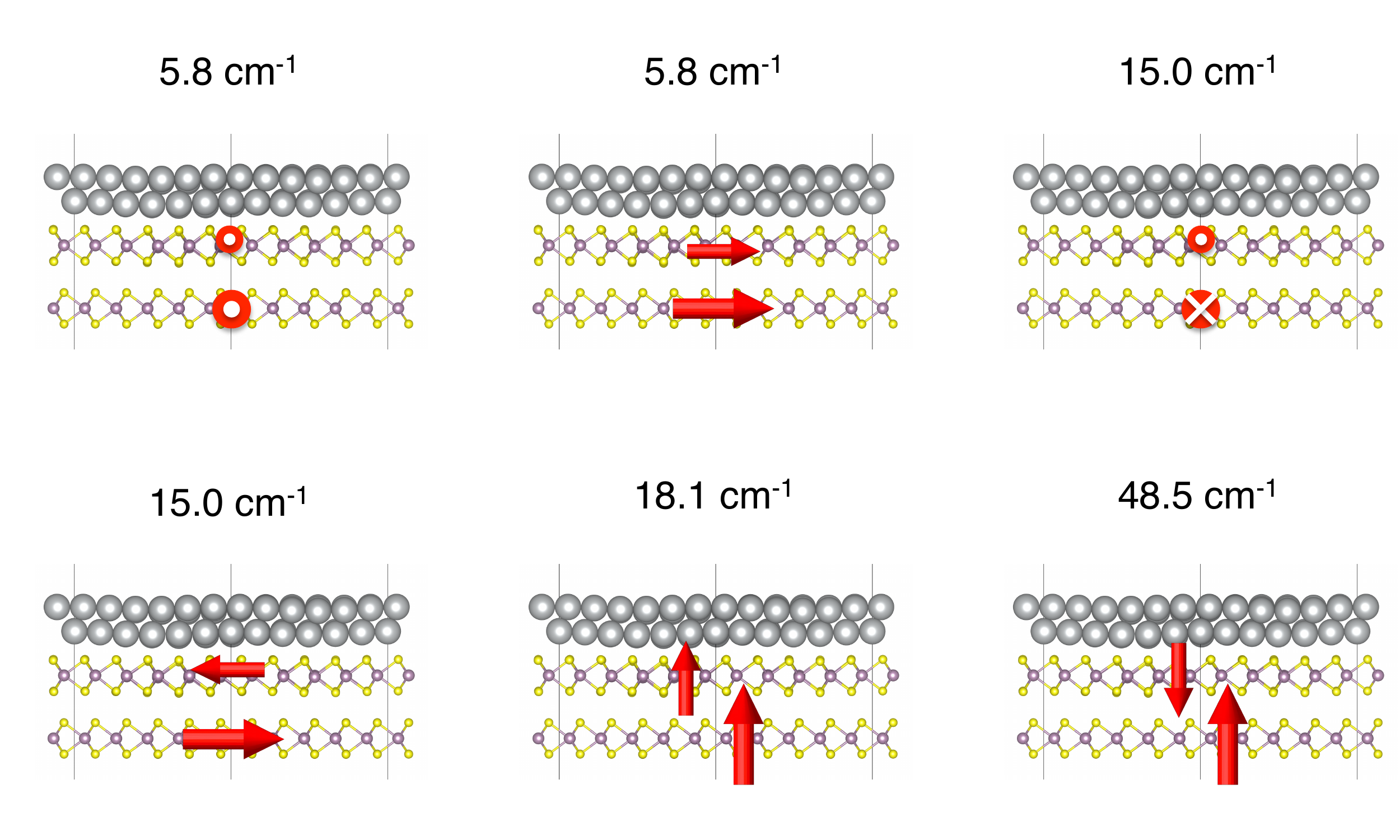}
\caption{Illustrations of the low-frequency interlayer modes' vibrations. Calculated harmonic frequencies at $\Gamma$-point are indicated above each vibration. The red arrows indicate the direction of the layer displacement in each shear (first four figures) and breathing (last two figures) mode, with crosses and circles indicating vibrations into and out-of the page, respectively.}
\label{Fig:LF-modes}
\end{figure}

Calculating the vibrational modes of the entire system of MoS$_2$ on Ag(111) is computationally more demanding than for the Ag-intercalated MoS$_2$ in the previous section~\ref{sec:results}. This is due to the substantially larger number of atoms comprising the $(6\times 6)$ Ag(111) layer (72 additional Ag atoms on top of the 150 atoms of the $(5\times 5)$ MoS$_2$ bilayer, which would require 708 separate DFT calculations to compute the vibrational properties of the composite structure within the FDM). Furthermore, the diagonalization of the full dynamical matrix would also yield many additional modes with vibrations primarily confined to the Ag atoms within the Ag(111) membrane while the MoS$_2$ atoms remain motionless. These substrate modes are of no significant relevance as we wish to focus on the effects on the Raman signature of MoS$_2$. Therefore, to reduce the computational cost and to obtain only the relevant information on the phonon modes of MoS$_2$ on Ag(111), we employed the quasiparticle finite-displacement (QPFD) method as described in Ref.~\onlinecite{tristant2018finite}. In this method, the Ag film is treated as a single quasiparticle, which allows for all Ag atoms to be rigidly displaced by the same amount in a given direction to compute the inter-atomic forces. The modified dynamical matrix was then constructed from these forces. This approach is an approximation that relies on the existence of low-frequency modes where the silver layers move quasi-rigidly. The validity of the approximation can be measured by the absence of complex frequencies in the resulting quasiparticle phonon bandstructures as can be seen in Fig. S14 [link to SI]. This approach reduces the number of calculations by a factor of two and facilitates the analysis of the results. Diagonalizing the modified dynamical matrix gives access to vibrational modes which only include the vibrations within the bilayer MoS$_2$ mediated by the interaction with Ag(111) and the vibrations at the interface between MoS$_2$ and Ag(111).

Fig.~\ref{Fig:LF-modes} shows the interfacial LF interlayer modes of MoS$_2$ placed on Ag(111). There are two doubly degenerate shear modes and two breathing modes at the $\Gamma$-point with symmetries allowing for Raman scattering. The first SM corresponds to a in-phase lateral vibration of the two MoS$_2$ layers with respect to the Ag-film and has a phonon frequency of about 6 cm$^{-1}$. This is a new mode that would be absent without the Ag membrane. The small frequency of this mode would place the corresponding Raman peak very close to the strong Rayleigh line in an experimental spectrum making the observation of this mode quite challenging. The next SM shows out-of-phase lateral MoS$_2$-layer vibrations and has a larger phonon frequency of about 15 cm$^{-1}$. This mode can be viewed as a modified version of the original SM in pure 2L MoS$_2$, whose frequency is now red-shifted from its original value of about 25 cm$^{-1}$ due to Ag(111). Similarly, the first interfacial breathing mode is a new mode with in-phase vertical displacements of the two MoS$_2$ layers against the silver membrane and a phonon frequency of about 18~cm$^{-1}$. The corresponding Raman peak would be expected to be located in close proximity to the one of the second SM mode. The second LBM is characterized by out-of-phase vertical displacements of the two MoS$_2$ layers with a phonon frequency of 48.5~cm$^{-1}$. This mode corresponds to the modified version of the LBM in pristine bilayer MoS$_2$ and is now blue-shifted from the original value of about 39.5~cm$^{-1}$. Interestingly, the frequency shifts for the modified SM and LBM have almost the same magnitude, but occur in opposite directions. Note that when viewing the corresponding animations of these vibrations it can be seen that the displacement of the MoS$_2$ layer directly in contact with the Ag(111) film has a lower amplitude of motion than the second MoS$_2$ layer. This is due to a quite strong adhesion between Ag(111) and MoS$_2$ resulting in the observed frequency shifts of the interlayer modes after adsorption. Based on these theoretical results, the adsorption of an Ag-film is expected to strongly modify the vibrational properties of bilayer MoS$_2$. More importantly, the detection of the interfacial modes would confirm, experimentally, the creation of a low-energy and high-quality interface between the two materials. 

\subsection{\label{sec:limits}Comparison with Experiment}
As demonstrated using a DFT analysis, the renormalization of the LF phonons due to Ag-intercalation could be used for the determination of the silver-concentration in few-layer MoS$_2$ devices. Naturally, there are some limitations to directly transferring the results of the computational predictions to experiments. We will now discuss those limitations in some details, using evidence from published data (HF modes) and new experimental data focusing on LF modes that is presented here for the first time. 
\subsubsection{High-Frequency modes}
First, we will address the comparison to the previously reported experimental results on the HF modes by Domask~\cite{Domask2017}. There, samples with thickness of 6~nm, 9~nm, and 79~nm were characterized using Raman spectroscopy at three main sample preparation steps: 1) on exfoliation of the pure MoS$_2$ film, 2) on deposition of the Ag-film on top of MoS$_2$, and 3) after annealing for 4~hours at 400$^{\circ}$C at which point Ag-intercalation is assumed to have taken place. Between the first and the last steps, the positions of the HF Raman peaks are found to red-shift slightly in the two thinner samples while for the thick sample, no significant shift was recorded. The red-shifts in the thinner samples are below 1 cm$^{-1}$ with larger shifts for the $E_g$ mode compared to the $A_{1g}$ mode. 

Unfortunately, direct simulation of the experimental sample thickness is not feasible with DFT due to high computational cost, while LCM only applies to the LF interlayer modes, since it does not capture the more complicated intralayer force-constants. Thus, we can only speculate about the mechanisms taking place in the experiments of Domask~\cite{Domask2017} by examining trends of our HF-modes results in few-layer systems with increasing number of layers. As mentioned in section~\ref{sec:results}, the HF modes red-shift with $C_{\mathrm{Ag}}$ similar to the LF modes, but by a significantly smaller amount. In fact, the frequency of the $E_g$ mode remains practically constant up to $C_{\mathrm{Ag}}=1\,$at\% for all numbers of layers. For the $A_{1g}$ mode's frequency, the red-shift rates up to $C_{\mathrm{Ag}}=1\,$at\% are about -3.6~cm$^{-1}$/at\%, -1.6~cm$^{-1}$/at\%, and -1.4~cm$^{-1}$/at\% for 2L, 3L, and 4L, respectively. The observation that the $A_{1g}$ mode experiences a stronger red-shift for increasing $C_{\mathrm{Ag}}$ compared to the $E_g$ mode is consistent with the previously reported renormalization of the same modes in single-layer MoS$_2$ samples with increasing carrier concentration~\cite{Chakraborty2012}. There, the $A_{1g}$ mode has been reported to red-shift significantly with carrier concentration (up to 4 cm$^{-1}$) while the $E_g$ (labeled $E^{1}_{2g}$ in monolayer) showed only a very weak red-shift (maximum shift below 1~cm$^{-1}$). These contrasting trends have been explained based on the symmetry of the modes and how the symmetry affects the underlying electron-phonon coupling (EPC), leading to an increasing EPC with carrier concentration for the $A_{1g}$ mode while for the $E_g$ it is practically vanishing throughout the carrier concentration range. The demonstrated renormalization of the $A_{1g}$ mode has been proposed as a readout of the carrier concentration in MoS$_2$ devices~\cite{Chakraborty2012}. While we did not examine the EPC here, the symmetry argument is nevertheless useful for the understanding of the present results, as well. In the $E_g$ mode, the sulfur atoms vibrate in the in-plane direction, while in the $A_{1g}$ mode they are displaced in the out-of-plane direction~\cite{Terrones2014}. Similar to the discussion for the LF modes, the in-plane structural features of MoS$_2$ are practically unaffected by Ag-impurities, but the out-of-plane ones change dramatically after silver intercalation resulting in the significant red-shifts of those phonon frequencies whose vibrations involve out-of-plane atomic motions. Thus, the experimental findings of Domask are somewhat counter-intuitive. One simple explanation could be that experimentally silver diffused only into a few vdW gaps of the MoS$_2$ and together with the relatively large sample thickness, the expected red-shifts are averaged out over the large number of gaps not containing Ag consistent with the diminishing $\partial \omega/ \partial C_{\mathrm{Ag}}$ slopes with increasing number of layers observed in our calculations. 

\subsubsection{Low-Frequency modes}
The theoretical analysis developed in this paper is primarily focused on the LF modes. We demonstrated that LF modes are more sensitive to Ag-intercalation into the vdW gaps of MoS$_2$ than their HF counterparts. To date, the LF modes have not been studied experimentally as systematically as the HF ones due to challenges in their experimental detection. To compare predictions with experiments, the LF  Raman response in few-layer MoS$_2$ has been recorded experimentally and is reported here for the first time.

As outlined in section II B, silver was first deposited on bulk MoS$_2$, followed by annealing to allow for Ag-diffusion into the bulk, and finally, Ag-intercalated MoS$_2$ flakes were mechanically exfoliated. Out of the thousands of flakes exfoliated from the same bulk on a substrate, few-layer \textit{vs.} thicker samples were distinguished based on the known change in optical contrast with thickness \cite{Li2012, Li2013}. The few-layer selection was then characterized using Raman spectroscopy. Examples of optical micrograph images together with the associated Raman spectra are shown in Fig. S17 [link to SI]. Interestingly, the Raman response of most of these flakes matched that of pure 2L and 3L MoS$_2$ samples \cite{Li2012,Liang2017}. However, one flake (hereafter referred to as flake d), while optically appearing like a 2L sample, exhibited a strong change of its LF Raman spectrum that could not be ascribed to pure MoS$_2$ of any thickness or polytype \cite{Liang2017}. At the same time, the shifts of the LF modes' positions for flake d are consistent with our DFT predictions for 2L Ag-intercalated MoS$_2$. We note that using optical contrast alone results in some uncertainty when determining the experimental number of layers, and we consider details of the Raman spectra to further evaluate our assumptions. Future studies could include atomic force microscopy to support sample-thickness assignment. 

\begin{figure}[ht!]
\centering
\includegraphics[width=0.4\textwidth,keepaspectratio]
{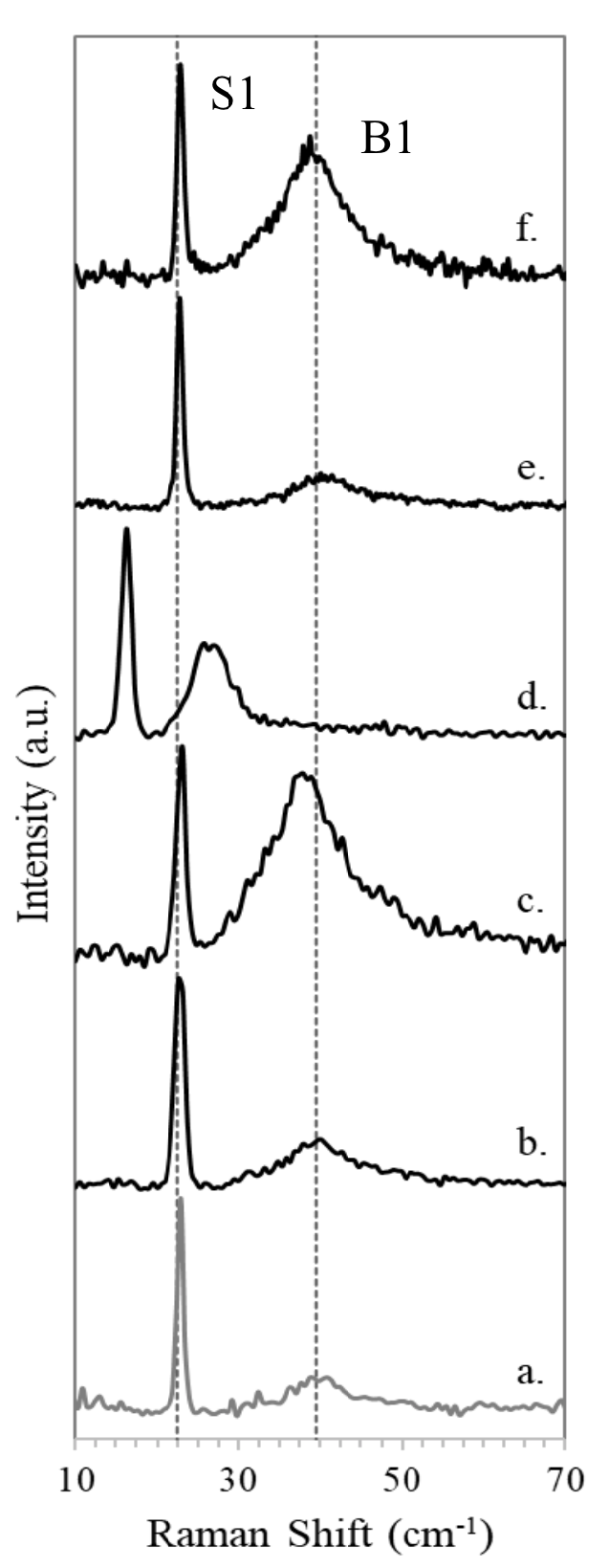}
\caption{\label{fig:2L_exp}LF Raman spectra of 2L MoS$_2$ flakes. Flake a is from the crystal before it was annealed with the Ag film on it. Flakes b–d are from crystals annealed at 400$^{\circ}$C for 24 hours after Ag deposition, and flakes e–f are from crystals annealed at 500$^{\circ}$C for 24 hours after Ag deposition}
\end{figure}

Raman spectra were collected from twelve suspected 2L flakes, and spectra from six flakes are shown in Fig.~\ref{fig:2L_exp} with peak positions provided in table~\ref{tab:exp_peaks}. The remaining six flakes not shown all matched pure 2L MoS$_2$. Three flakes (c, d, and f) showed significant spectral changes, and their LF spectra are displayed along with a spectrum from the crystal prior to Ag deposition (flake a), a representative flake without changes after annealing at 400$^{\circ}$C (flake b), and a representative flake without changes after annealing at 500$^{\circ}$C (flake e). Flake d showed large red-shifts of both of the LF modes and was obtained from the crystal annealed at 400$^{\circ}$C after Ag deposition. The SM mode red-shifted by 6.61~cm$^{-1}$ and the LBM mode by 13.42~cm$^{-1}$. These red-shifts are consistent with 0.9 (LBM)--1.2 (SM) at.\% Ag based on their positions, suggesting significant intercalation. The intensity of the LBM relative to the SM (ILBM/ISM) from flake d (0.48) is also slightly greater than it is from the pristine flake a (0.25). Spectra from flakes c (crystal annealed at 400$^{\circ}$C) and f (annealed at 500$^{\circ}$C) exhibited only minor red-shifts of the LBM of 2.19 and 1.02~cm$^{-1}$, respectively, and no significant changes to the position of the SM.

\begin{table}[ht!]
\centering
\begin{ruledtabular}
\begin{tabular}{ccc}
Flake & SM [cm-1] & LBM [cm-1] \\
a & 22.87 & 40.08 \\
b & 22.8 & 40.09 \\
c & 22.94 & 37.89 \\
d & 16.26 & 26.66 \\
e & 22.75 & 40.64 \\
f & 22.93 & 39.07
\end{tabular}
\end{ruledtabular}
\caption{\label{tab:exp_peaks}Peak positions for the Raman spectra from six 2L MoS$_2$ flakes. The spectra are shown in Fig. (new experimental figure).}
\end{table}

\begin{figure}[ht!]
\centering
\includegraphics[width=0.3\textwidth,keepaspectratio]
{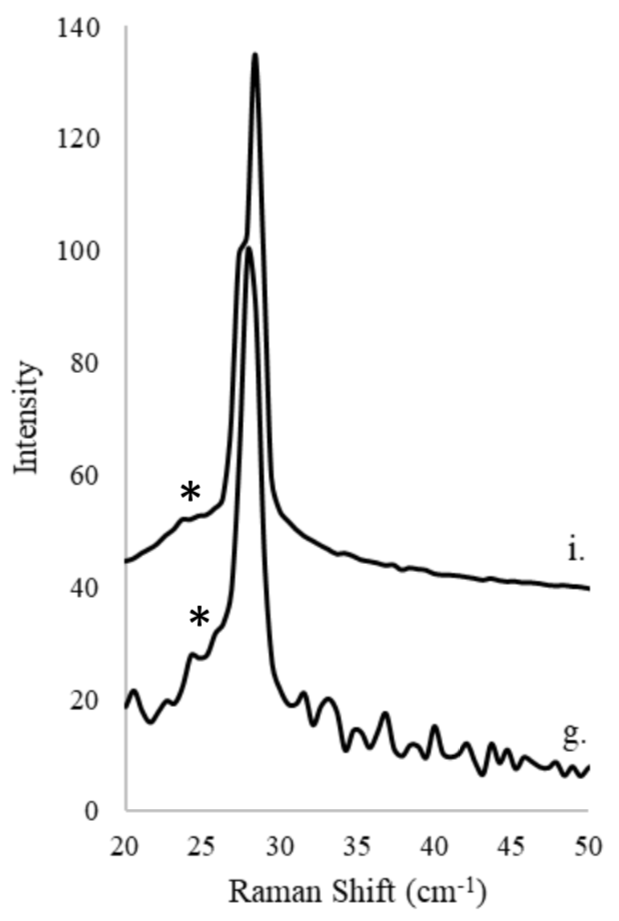}
\caption{\label{fig:3L_exp_shoulders}Raman spectra from 3L MoS$_2$ showing a weak shoulder (*) between 24 cm$^{-1}$ and 25 cm$^{-1}$. Spectra from all 3L flakes (a–i) are included in the SI. Flake g was annealed with Ag at 400$^{\circ}$C and flake i at 500$^{\circ}$C. }
\end{figure}

Further, Raman spectra from nine suspected 3L flakes were collected, with details and spectra provided in Fig. S18 [link to SI]. The positions of the LBM and SM are close enough to each other that they appear as a single peak~\cite{Zhang2013}. We found few changes that could not be otherwise explained, but we did observe weak, broad shoulders between 24 and 25 cm$^{-1}$ on the peak at 28 cm$^{-1}$ from two flakes, as shown in Fig.~\ref{fig:3L_exp_shoulders}. Note that the narrow shoulder very close to the peak at 28 cm$^{-1}$ from one of the flakes is not mirrored in the anti-Stokes spectrum and is believed to be an artifact. We could not ascribe the broad shoulders to any other few-layer thicknesses or stacking order~\cite{Xia2017}. Table~\ref{tab:app_chap4_slopesallL} shows that the LBM is predicted to red-shift more than the SM due to intercalation of Ag, so we speculate that splitting of the peaks could be due to a red-shift of the LBM by approximately 3.5 cm$^{-1}$. However, the SM did not shift compared to reference spectra from pristine 3L MoS$_2$, even though a modest red-shift was predicted. Regarding this discrepancy, we note that silver-intercalated 3L MoS$_2$ was also modeled with Ag atoms in both vdW gaps (see figures S3 and S4 in SI [link to SI], resulting in an even stronger red-shift of both LF modes compared to Ag in just one of the gaps. The experimental absence of a shift for the SM is likely stemming from aspects that are not taken into account in the theoretical analysis, such as clustering, defects, and other structural features. 

The above discussions of the experimental findings highlight the challenges of experimental characterization. While computational simulations allow for exact control of one specific variable of the system, experimental samples can have multiple features being modified at the same time leading to an overall ambiguous signal. Furthermore, in section~\ref{subsec:linchain} we demonstrated that the variation of the number of layers in thick samples, the number of vdW gaps which Ag-impurities occupy, and the silver-concentration within the gaps can all lead to similar Raman shifts, especially for small silver concentrations. The present discussion was also based on the assumption that silver is distributed evenly throughout the lateral extensions of the vdW gaps as simulated by periodic boundary conditions. However, an experimental study of Ag-metal contacts on monolayer MoS$_2$ has reported inhomogeneous silver-distribution and formation of Ag-islands of different shapes and sizes~\cite{Gong2013}. Analogous Ag-agglomeration is conceivable in few-layer MoS$_2$, as well. Simulations of localized Ag-islands are currently intractable within DFT and would require the use of other techniques such as molecular dynamics simulations, which are beyond the scope of the present study.

Finally, in this study we have not computationally addressed the actual Raman intensities because the corresponding computations were not feasible for the large supercells required for the present simulations. Naturally, Ag-intercalation is likely to affect Raman intensity profiles of MoS$_2$. For example, in some of our experiments we have observed a quenching of the Raman signal when a thick film of silver was deposited on few-layer MoS$_2$ and no annealing was performed. In general, low Raman intensity might present an additional experimental challenge for the detection and usage of LF Raman signature of silver-intercalated MoS$_2$ devices.

\section{\label{sec:conclusion}Conclusion}
We have systematically studied the vibrational properties of few-layer MoS$_2$ intercalated with silver with a specific focus on the LF interlayer modes. The LF interlayer modes are significantly more sensitive to Ag-intercalation compared to their HF intralayer counterparts. This makes them attractive not only as a practical gauge for determination of silver-concentration, but also more generally as a platform to study how the interlayer interactions are modified by impurities in the vdW gap. In this way, the LF modes provide valuable insights that are possibly outweighed by the challenges of their experimental detection. Thus, the experimental measurement of LF interlayer modes should be included as part of the routine Raman characterization step in addition to the commonly recorded HF intralayer modes.

Several examples from our experimental work suggest intercalation of Ag into few-layer MoS$_2$ upon annealing a crystal of MoS$_2$ with a Ag film on it at 400$^{\circ}$C or 500$^{\circ}$C, although we also found many examples of flakes with LF Raman spectra that were unaffected by diffusion of Ag, possibly due to inhomogeneity of the sample.

Possible limitations of unambiguous $C_{\mathrm{Ag}}$ determination in realistic samples have been discussed in connection with experiments. Multiple aspects, like the number of layers in thick samples or the number of vdW gaps containing Ag-impurities, have been theoretically shown to result in very similar Raman signatures. Together with additional effects potentially present in realistic samples decoding of all the different effects only from Raman measurements might prove challenging in practice. Nevertheless, Raman spectroscopy could potentially be used in conjunction with other structural characterization techniques to unambiguously determine the Ag-impurity concentration in few-layer MoS$_2$. At the minimum, Raman spectroscopy can be used to determine successful intercalation, even if determining the actual concentration, while feasible in principle, requires very high-resolution measurements and control over sample parameters. 

Finally, we expect that the discussion provided here applies to the cases of other metal atoms intercalated in few-layer MoS$_2$ or even in other TMDCs of similar crystal structure, so long at the atomic radius of the impurities can be accommodated in the vdW gap of the few-layer samples. Thus, the groundwork laid here for the metal-intercalation detection with Raman spectroscopy might find broader applications in the context of the many other metals that have been explored experimentally for contact improvement to MoS$_2$ ~\cite{Freedy2019,Freedy2020,Walter2020}. The actual frequency shifts are material dependent but the fundamental mechanisms described herein are expected to hold true. 


\section{Acknowledgements}
Most of this work was performed using supercomputing resources provided by the Center for Computational Innovations (CCI) at Rensselaer Polytechnic Institute. L.L. acknowledges work conducted at the Center for Nanophase Materials Sciences, which is a US Department of Energy Office of Science User Facility. Laboratory experiments were performed at Penn State with support from NSF through the EFRI 2-DARE 1433378 Research Experiences and Mentoring supplement. L.K. and S.M. are grateful to Ama D. Agyapong for depositing Ag films and Maxwell Wetherington in Penn State’s Materials Characterization Laboratory for training on Raman spectroscopy.
\bibliography{references,methods}

\begin{thebibliography}{42}%
\makeatletter
\providecommand \@ifxundefined [1]{%
 \@ifx{#1\undefined}
}%
\providecommand \@ifnum [1]{%
 \ifnum #1\expandafter \@firstoftwo
 \else \expandafter \@secondoftwo
 \fi
}%
\providecommand \@ifx [1]{%
 \ifx #1\expandafter \@firstoftwo
 \else \expandafter \@secondoftwo
 \fi
}%
\providecommand \natexlab [1]{#1}%
\providecommand \enquote  [1]{``#1''}%
\providecommand \bibnamefont  [1]{#1}%
\providecommand \bibfnamefont [1]{#1}%
\providecommand \citenamefont [1]{#1}%
\providecommand \href@noop [0]{\@secondoftwo}%
\providecommand \href [0]{\begingroup \@sanitize@url \@href}%
\providecommand \@href[1]{\@@startlink{#1}\@@href}%
\providecommand \@@href[1]{\endgroup#1\@@endlink}%
\providecommand \@sanitize@url [0]{\catcode `\\12\catcode `\$12\catcode
  `\&12\catcode `\#12\catcode `\^12\catcode `\_12\catcode `\%12\relax}%
\providecommand \@@startlink[1]{}%
\providecommand \@@endlink[0]{}%
\providecommand \url  [0]{\begingroup\@sanitize@url \@url }%
\providecommand \@url [1]{\endgroup\@href {#1}{\urlprefix }}%
\providecommand \urlprefix  [0]{URL }%
\providecommand \Eprint [0]{\href }%
\providecommand \doibase [0]{http://dx.doi.org/}%
\providecommand \selectlanguage [0]{\@gobble}%
\providecommand \bibinfo  [0]{\@secondoftwo}%
\providecommand \bibfield  [0]{\@secondoftwo}%
\providecommand \translation [1]{[#1]}%
\providecommand \BibitemOpen [0]{}%
\providecommand \bibitemStop [0]{}%
\providecommand \bibitemNoStop [0]{.\EOS\space}%
\providecommand \EOS [0]{\spacefactor3000\relax}%
\providecommand \BibitemShut  [1]{\csname bibitem#1\endcsname}%
\let\auto@bib@innerbib\@empty
\bibitem [{\citenamefont {Radisavljevic}\ \emph {et~al.}(2011)\citenamefont
  {Radisavljevic}, \citenamefont {Radenovic}, \citenamefont {Brivio},
  \citenamefont {Giacometti},\ and\ \citenamefont {Kis}}]{Radisavljevic2011}%
  \BibitemOpen
  \bibfield  {author} {\bibinfo {author} {\bibfnamefont {B.}~\bibnamefont
  {Radisavljevic}}, \bibinfo {author} {\bibfnamefont {A.}~\bibnamefont
  {Radenovic}}, \bibinfo {author} {\bibfnamefont {J.}~\bibnamefont {Brivio}},
  \bibinfo {author} {\bibfnamefont {V.}~\bibnamefont {Giacometti}}, \ and\
  \bibinfo {author} {\bibfnamefont {A.}~\bibnamefont {Kis}},\ }\href@noop {}
  {\bibfield  {journal} {\bibinfo  {journal} {Nature Nanotechnology}\ }\textbf
  {\bibinfo {volume} {6}},\ \bibinfo {pages} {147} (\bibinfo {year}
  {2011})}\BibitemShut {NoStop}%
\bibitem [{\citenamefont {{Majumdar}}\ \emph {et~al.}(2014)\citenamefont
  {{Majumdar}}, \citenamefont {{Hobbs}},\ and\ \citenamefont
  {{Kirsch}}}]{Majumdar2014}%
  \BibitemOpen
  \bibfield  {author} {\bibinfo {author} {\bibfnamefont {K.}~\bibnamefont
  {{Majumdar}}}, \bibinfo {author} {\bibfnamefont {C.}~\bibnamefont {{Hobbs}}},
  \ and\ \bibinfo {author} {\bibfnamefont {P.~D.}\ \bibnamefont {{Kirsch}}},\
  }\href@noop {} {\bibfield  {journal} {\bibinfo  {journal} {IEEE Electron
  Device Letters}\ }\textbf {\bibinfo {volume} {35}},\ \bibinfo {pages} {402}
  (\bibinfo {year} {2014})}\BibitemShut {NoStop}%
\bibitem [{\citenamefont {{Kuzum}}\ \emph {et~al.}(2011)\citenamefont
  {{Kuzum}}, \citenamefont {{Krishnamohan}}, \citenamefont {{Nainani}},
  \citenamefont {{Sun}}, \citenamefont {{Pianetta}}, \citenamefont {{Wong}},\
  and\ \citenamefont {{Saraswat}}}]{Kuzum2011}%
  \BibitemOpen
  \bibfield  {author} {\bibinfo {author} {\bibfnamefont {D.}~\bibnamefont
  {{Kuzum}}}, \bibinfo {author} {\bibfnamefont {T.}~\bibnamefont
  {{Krishnamohan}}}, \bibinfo {author} {\bibfnamefont {A.}~\bibnamefont
  {{Nainani}}}, \bibinfo {author} {\bibfnamefont {Y.}~\bibnamefont {{Sun}}},
  \bibinfo {author} {\bibfnamefont {P.~A.}\ \bibnamefont {{Pianetta}}},
  \bibinfo {author} {\bibfnamefont {H.-S.~P.}\ \bibnamefont {{Wong}}}, \ and\
  \bibinfo {author} {\bibfnamefont {K.~C.}\ \bibnamefont {{Saraswat}}},\
  }\href@noop {} {\bibfield  {journal} {\bibinfo  {journal} {IEEE Trans.
  Electron Devices}\ }\textbf {\bibinfo {volume} {58}},\ \bibinfo {pages} {59}
  (\bibinfo {year} {2011})}\BibitemShut {NoStop}%
\bibitem [{\citenamefont {Rai}\ \emph {et~al.}(2018)\citenamefont {Rai},
  \citenamefont {Movva}, \citenamefont {Roy}, \citenamefont {Taneja},
  \citenamefont {Chowdhury},\ and\ \citenamefont {Banerjee}}]{Rai2018}%
  \BibitemOpen
  \bibfield  {author} {\bibinfo {author} {\bibfnamefont {A.}~\bibnamefont
  {Rai}}, \bibinfo {author} {\bibfnamefont {H.}~\bibnamefont {Movva}}, \bibinfo
  {author} {\bibfnamefont {A.}~\bibnamefont {Roy}}, \bibinfo {author}
  {\bibfnamefont {D.}~\bibnamefont {Taneja}}, \bibinfo {author} {\bibfnamefont
  {S.}~\bibnamefont {Chowdhury}}, \ and\ \bibinfo {author} {\bibfnamefont
  {S.}~\bibnamefont {Banerjee}},\ }\href {\doibase 10.3390/cryst8080316}
  {\bibfield  {journal} {\bibinfo  {journal} {Crystals}\ }\textbf {\bibinfo
  {volume} {8}},\ \bibinfo {pages} {316} (\bibinfo {year} {2018})}\BibitemShut
  {NoStop}%
\bibitem [{\citenamefont {Schulman}\ \emph {et~al.}(2018)\citenamefont
  {Schulman}, \citenamefont {Arnold},\ and\ \citenamefont
  {Das}}]{Schulman2018}%
  \BibitemOpen
  \bibfield  {author} {\bibinfo {author} {\bibfnamefont {D.~S.}\ \bibnamefont
  {Schulman}}, \bibinfo {author} {\bibfnamefont {A.~J.}\ \bibnamefont
  {Arnold}}, \ and\ \bibinfo {author} {\bibfnamefont {S.}~\bibnamefont {Das}},\
  }\href {\doibase 10.1039/C7CS00828G} {\bibfield  {journal} {\bibinfo
  {journal} {Chem. Soc. Rev.}\ }\textbf {\bibinfo {volume} {47}},\ \bibinfo
  {pages} {3037} (\bibinfo {year} {2018})}\BibitemShut {NoStop}%
\bibitem [{\citenamefont {Freedy}\ and\ \citenamefont
  {McDonnell}(2020)}]{Freedy2020}%
  \BibitemOpen
  \bibfield  {author} {\bibinfo {author} {\bibfnamefont {K.~M.}\ \bibnamefont
  {Freedy}}\ and\ \bibinfo {author} {\bibfnamefont {S.~J.}\ \bibnamefont
  {McDonnell}},\ }\href {\doibase 10.3390/ma13030693} {\enquote {\bibinfo
  {title} {Contacts for molybdenum disulfide: Interface chemistry and thermal
  stability},}\ } (\bibinfo {year} {2020})\BibitemShut {NoStop}%
\bibitem [{\citenamefont {Abraham}\ and\ \citenamefont
  {Mohney}(2017)}]{Abraham2017}%
  \BibitemOpen
  \bibfield  {author} {\bibinfo {author} {\bibfnamefont {M.}~\bibnamefont
  {Abraham}}\ and\ \bibinfo {author} {\bibfnamefont {S.~E.}\ \bibnamefont
  {Mohney}},\ }\href@noop {} {\bibfield  {journal} {\bibinfo  {journal}
  {Journal of Applied Physics}\ }\textbf {\bibinfo {volume} {122}},\ \bibinfo
  {pages} {115306} (\bibinfo {year} {2017})}\BibitemShut {NoStop}%
\bibitem [{\citenamefont {Souder}\ and\ \citenamefont
  {Brodie}(1971)}]{Souder1971}%
  \BibitemOpen
  \bibfield  {author} {\bibinfo {author} {\bibfnamefont {A.~D.}\ \bibnamefont
  {Souder}}\ and\ \bibinfo {author} {\bibfnamefont {D.~E.}\ \bibnamefont
  {Brodie}},\ }\href@noop {} {\bibfield  {journal} {\bibinfo  {journal}
  {Canadian Journal of Physics}\ }\textbf {\bibinfo {volume} {49}},\ \bibinfo
  {pages} {2565} (\bibinfo {year} {1971})}\BibitemShut {NoStop}%
\bibitem [{\citenamefont {Souder}\ and\ \citenamefont
  {Brodie}(1972)}]{Souder1972}%
  \BibitemOpen
  \bibfield  {author} {\bibinfo {author} {\bibfnamefont {A.~D.}\ \bibnamefont
  {Souder}}\ and\ \bibinfo {author} {\bibfnamefont {D.~E.}\ \bibnamefont
  {Brodie}},\ }\href@noop {} {\bibfield  {journal} {\bibinfo  {journal}
  {Canadian Journal of Physics}\ }\textbf {\bibinfo {volume} {50}},\ \bibinfo
  {pages} {1223} (\bibinfo {year} {1972})}\BibitemShut {NoStop}%
\bibitem [{\citenamefont {Liang}\ \emph
  {et~al.}(2017{\natexlab{a}})\citenamefont {Liang}, \citenamefont {Zhang},
  \citenamefont {Sumpter}, \citenamefont {Tan}, \citenamefont {Tan},\ and\
  \citenamefont {Meunier}}]{Liang2017}%
  \BibitemOpen
  \bibfield  {author} {\bibinfo {author} {\bibfnamefont {L.}~\bibnamefont
  {Liang}}, \bibinfo {author} {\bibfnamefont {J.}~\bibnamefont {Zhang}},
  \bibinfo {author} {\bibfnamefont {B.~G.}\ \bibnamefont {Sumpter}}, \bibinfo
  {author} {\bibfnamefont {Q.~H.}\ \bibnamefont {Tan}}, \bibinfo {author}
  {\bibfnamefont {P.~H.}\ \bibnamefont {Tan}}, \ and\ \bibinfo {author}
  {\bibfnamefont {V.}~\bibnamefont {Meunier}},\ }\href@noop {} {\bibfield
  {journal} {\bibinfo  {journal} {ACS Nano}\ }\textbf {\bibinfo {volume}
  {11}},\ \bibinfo {pages} {11777} (\bibinfo {year}
  {2017}{\natexlab{a}})}\BibitemShut {NoStop}%
\bibitem [{\citenamefont {Rice}\ \emph {et~al.}(2013)\citenamefont {Rice},
  \citenamefont {Young}, \citenamefont {Zan}, \citenamefont {Bangert},
  \citenamefont {Wolverson}, \citenamefont {Georgiou}, \citenamefont {Jalil},\
  and\ \citenamefont {Novoselov}}]{Rice2013}%
  \BibitemOpen
  \bibfield  {author} {\bibinfo {author} {\bibfnamefont {C.}~\bibnamefont
  {Rice}}, \bibinfo {author} {\bibfnamefont {R.~J.}\ \bibnamefont {Young}},
  \bibinfo {author} {\bibfnamefont {R.}~\bibnamefont {Zan}}, \bibinfo {author}
  {\bibfnamefont {U.}~\bibnamefont {Bangert}}, \bibinfo {author} {\bibfnamefont
  {D.}~\bibnamefont {Wolverson}}, \bibinfo {author} {\bibfnamefont
  {T.}~\bibnamefont {Georgiou}}, \bibinfo {author} {\bibfnamefont
  {R.}~\bibnamefont {Jalil}}, \ and\ \bibinfo {author} {\bibfnamefont {K.~S.}\
  \bibnamefont {Novoselov}},\ }\href@noop {} {\bibfield  {journal} {\bibinfo
  {journal} {Phys. Rev. B}\ }\textbf {\bibinfo {volume} {87}},\ \bibinfo
  {pages} {081307} (\bibinfo {year} {2013})}\BibitemShut {NoStop}%
\bibitem [{\citenamefont {Parkin}\ \emph {et~al.}(2016)\citenamefont {Parkin},
  \citenamefont {Balan}, \citenamefont {Liang}, \citenamefont {Das},
  \citenamefont {Lamparski}, \citenamefont {Naylor}, \citenamefont
  {Rodríguez-Manzo}, \citenamefont {Johnson}, \citenamefont {Meunier},\ and\
  \citenamefont {Drndić}}]{Parkin2016}%
  \BibitemOpen
  \bibfield  {author} {\bibinfo {author} {\bibfnamefont {W.~M.}\ \bibnamefont
  {Parkin}}, \bibinfo {author} {\bibfnamefont {A.}~\bibnamefont {Balan}},
  \bibinfo {author} {\bibfnamefont {L.}~\bibnamefont {Liang}}, \bibinfo
  {author} {\bibfnamefont {P.~M.}\ \bibnamefont {Das}}, \bibinfo {author}
  {\bibfnamefont {M.}~\bibnamefont {Lamparski}}, \bibinfo {author}
  {\bibfnamefont {C.~H.}\ \bibnamefont {Naylor}}, \bibinfo {author}
  {\bibfnamefont {J.~A.}\ \bibnamefont {Rodríguez-Manzo}}, \bibinfo {author}
  {\bibfnamefont {A.~T.~C.}\ \bibnamefont {Johnson}}, \bibinfo {author}
  {\bibfnamefont {V.}~\bibnamefont {Meunier}}, \ and\ \bibinfo {author}
  {\bibfnamefont {M.}~\bibnamefont {Drndić}},\ }\href@noop {} {\bibfield
  {journal} {\bibinfo  {journal} {ACS Nano}\ }\textbf {\bibinfo {volume}
  {10}},\ \bibinfo {pages} {4134} (\bibinfo {year} {2016})}\BibitemShut
  {NoStop}%
\bibitem [{\citenamefont {Domask}(2017)}]{Domask2017}%
  \BibitemOpen
  \bibfield  {author} {\bibinfo {author} {\bibfnamefont {A.}~\bibnamefont
  {Domask}},\ }\emph {\bibinfo {title} {Reactivity and epitaxy of metal thin
  films on few-layer molybdenum disulfide}},\ \href@noop {} {Ph.D. thesis},\
  \bibinfo  {school} {The Pennsylvania State University} (\bibinfo {year}
  {2017})\BibitemShut {NoStop}%
\bibitem [{\citenamefont {Zhang}\ \emph {et~al.}(2016)\citenamefont {Zhang},
  \citenamefont {Tan}, \citenamefont {Wu}, \citenamefont {Shi},\ and\
  \citenamefont {Tan}}]{Zhang2016review}%
  \BibitemOpen
  \bibfield  {author} {\bibinfo {author} {\bibfnamefont {X.}~\bibnamefont
  {Zhang}}, \bibinfo {author} {\bibfnamefont {Q.-H.}\ \bibnamefont {Tan}},
  \bibinfo {author} {\bibfnamefont {J.-B.}\ \bibnamefont {Wu}}, \bibinfo
  {author} {\bibfnamefont {W.}~\bibnamefont {Shi}}, \ and\ \bibinfo {author}
  {\bibfnamefont {P.-H.}\ \bibnamefont {Tan}},\ }\href {\doibase
  10.1039/C5NR07205K} {\bibfield  {journal} {\bibinfo  {journal} {Nanoscale}\
  }\textbf {\bibinfo {volume} {8}},\ \bibinfo {pages} {6435} (\bibinfo {year}
  {2016})}\BibitemShut {NoStop}%
\bibitem [{\citenamefont {Kohn}\ and\ \citenamefont {Sham}(1965)}]{Kohn1965}%
  \BibitemOpen
  \bibfield  {author} {\bibinfo {author} {\bibfnamefont {W.}~\bibnamefont
  {Kohn}}\ and\ \bibinfo {author} {\bibfnamefont {L.~J.}\ \bibnamefont
  {Sham}},\ }\href {\doibase 10.1103/PhysRev.140.A1133} {\bibfield  {journal}
  {\bibinfo  {journal} {Phys. Rev.}\ }\textbf {\bibinfo {volume} {140}},\
  \bibinfo {pages} {A1133} (\bibinfo {year} {1965})}\BibitemShut {NoStop}%
\bibitem [{\citenamefont {Kresse}\ and\ \citenamefont
  {Furthm{\"{u}}ller}(1996)}]{Kresse1996}%
  \BibitemOpen
  \bibfield  {author} {\bibinfo {author} {\bibfnamefont {G.}~\bibnamefont
  {Kresse}}\ and\ \bibinfo {author} {\bibfnamefont {J.}~\bibnamefont
  {Furthm{\"{u}}ller}},\ }\href {\doibase 10.1103/PhysRevB.54.11169} {\bibfield
   {journal} {\bibinfo  {journal} {Phys. Rev. B}\ }\textbf {\bibinfo {volume}
  {54}},\ \bibinfo {pages} {11169} (\bibinfo {year} {1996})}\BibitemShut
  {NoStop}%
\bibitem [{\citenamefont {Kresse}\ and\ \citenamefont
  {Joubert}(1999)}]{Kresse1999}%
  \BibitemOpen
  \bibfield  {author} {\bibinfo {author} {\bibfnamefont {G.}~\bibnamefont
  {Kresse}}\ and\ \bibinfo {author} {\bibfnamefont {D.}~\bibnamefont
  {Joubert}},\ }\href {\doibase 10.1103/PhysRevB.59.1758} {\bibfield  {journal}
  {\bibinfo  {journal} {Physical Review B}\ }\textbf {\bibinfo {volume} {59}},\
  \bibinfo {pages} {1758} (\bibinfo {year} {1999})}\BibitemShut {NoStop}%
\bibitem [{\citenamefont {Bl{\"{o}}chl}(1994)}]{Blochl1994}%
  \BibitemOpen
  \bibfield  {author} {\bibinfo {author} {\bibfnamefont {P.~E.}\ \bibnamefont
  {Bl{\"{o}}chl}},\ }\href {\doibase 10.1103/PhysRevB.50.17953} {\bibfield
  {journal} {\bibinfo  {journal} {Phys. Rev. B}\ }\textbf {\bibinfo {volume}
  {50}},\ \bibinfo {pages} {17953} (\bibinfo {year} {1994})}\BibitemShut
  {NoStop}%
\bibitem [{\citenamefont {Monkhorst}\ and\ \citenamefont
  {Pack}(1976)}]{Monkhorst1976}%
  \BibitemOpen
  \bibfield  {author} {\bibinfo {author} {\bibfnamefont {H.~J.}\ \bibnamefont
  {Monkhorst}}\ and\ \bibinfo {author} {\bibfnamefont {J.~D.}\ \bibnamefont
  {Pack}},\ }\href@noop {} {\bibfield  {journal} {\bibinfo  {journal} {Phys.
  Rev. B}\ }\textbf {\bibinfo {volume} {13}},\ \bibinfo {pages} {5188}
  (\bibinfo {year} {1976})}\BibitemShut {NoStop}%
\bibitem [{\citenamefont {Freysoldt}\ \emph {et~al.}(2014)\citenamefont
  {Freysoldt}, \citenamefont {Grabowski}, \citenamefont {Hickel}, \citenamefont
  {Neugebauer}, \citenamefont {Kresse}, \citenamefont {Janotti},\ and\
  \citenamefont {{Van De Walle}}}]{Freysoldt2014}%
  \BibitemOpen
  \bibfield  {author} {\bibinfo {author} {\bibfnamefont {C.}~\bibnamefont
  {Freysoldt}}, \bibinfo {author} {\bibfnamefont {B.}~\bibnamefont
  {Grabowski}}, \bibinfo {author} {\bibfnamefont {T.}~\bibnamefont {Hickel}},
  \bibinfo {author} {\bibfnamefont {J.}~\bibnamefont {Neugebauer}}, \bibinfo
  {author} {\bibfnamefont {G.}~\bibnamefont {Kresse}}, \bibinfo {author}
  {\bibfnamefont {A.}~\bibnamefont {Janotti}}, \ and\ \bibinfo {author}
  {\bibfnamefont {C.~G.}\ \bibnamefont {{Van De Walle}}},\ }\href@noop {}
  {\bibfield  {journal} {\bibinfo  {journal} {Reviews of Modern Physics}\
  }\textbf {\bibinfo {volume} {86}},\ \bibinfo {pages} {253} (\bibinfo {year}
  {2014})}\BibitemShut {NoStop}%
\bibitem [{\citenamefont {Guzman}\ \emph {et~al.}(2017)\citenamefont {Guzman},
  \citenamefont {Onofrio},\ and\ \citenamefont {Strachan}}]{Guzman2017}%
  \BibitemOpen
  \bibfield  {author} {\bibinfo {author} {\bibfnamefont {D.~M.}\ \bibnamefont
  {Guzman}}, \bibinfo {author} {\bibfnamefont {N.}~\bibnamefont {Onofrio}}, \
  and\ \bibinfo {author} {\bibfnamefont {A.}~\bibnamefont {Strachan}},\
  }\href@noop {} {\bibfield  {journal} {\bibinfo  {journal} {Journal of Applied
  Physics}\ }\textbf {\bibinfo {volume} {121}},\ \bibinfo {pages} {055703}
  (\bibinfo {year} {2017})}\BibitemShut {NoStop}%
\bibitem [{\citenamefont {Togo}\ and\ \citenamefont {Tanaka}(2015)}]{Togo2015}%
  \BibitemOpen
  \bibfield  {author} {\bibinfo {author} {\bibfnamefont {A.}~\bibnamefont
  {Togo}}\ and\ \bibinfo {author} {\bibfnamefont {I.}~\bibnamefont {Tanaka}},\
  }\href {\doibase https://doi.org/10.1016/j.scriptamat.2015.07.021} {\bibfield
   {journal} {\bibinfo  {journal} {Scripta Materialia}\ }\textbf {\bibinfo
  {volume} {108}},\ \bibinfo {pages} {1} (\bibinfo {year} {2015})}\BibitemShut
  {NoStop}%
\bibitem [{\citenamefont {Li}\ \emph {et~al.}(2012)\citenamefont {Li},
  \citenamefont {Zhang}, \citenamefont {Yap}, \citenamefont {Tay},
  \citenamefont {Edwin}, \citenamefont {Olivier},\ and\ \citenamefont
  {Baillargeat}}]{Li2012}%
  \BibitemOpen
  \bibfield  {author} {\bibinfo {author} {\bibfnamefont {H.}~\bibnamefont
  {Li}}, \bibinfo {author} {\bibfnamefont {Q.}~\bibnamefont {Zhang}}, \bibinfo
  {author} {\bibfnamefont {C.~C.~R.}\ \bibnamefont {Yap}}, \bibinfo {author}
  {\bibfnamefont {B.~K.}\ \bibnamefont {Tay}}, \bibinfo {author} {\bibfnamefont
  {T.~H.~T.}\ \bibnamefont {Edwin}}, \bibinfo {author} {\bibfnamefont
  {A.}~\bibnamefont {Olivier}}, \ and\ \bibinfo {author} {\bibfnamefont
  {D.}~\bibnamefont {Baillargeat}},\ }\href {\doibase 10.1002/adfm.201102111}
  {\bibfield  {journal} {\bibinfo  {journal} {Advanced Functional Materials}\
  }\textbf {\bibinfo {volume} {22}},\ \bibinfo {pages} {1385} (\bibinfo {year}
  {2012})}\BibitemShut {NoStop}%
\bibitem [{\citenamefont {Li}\ \emph {et~al.}(2013)\citenamefont {Li},
  \citenamefont {Wu}, \citenamefont {Huang}, \citenamefont {Lu}, \citenamefont
  {Yang}, \citenamefont {Lu}, \citenamefont {Xiong},\ and\ \citenamefont
  {Zhang}}]{Li2013}%
  \BibitemOpen
  \bibfield  {author} {\bibinfo {author} {\bibfnamefont {H.}~\bibnamefont
  {Li}}, \bibinfo {author} {\bibfnamefont {J.}~\bibnamefont {Wu}}, \bibinfo
  {author} {\bibfnamefont {X.}~\bibnamefont {Huang}}, \bibinfo {author}
  {\bibfnamefont {G.}~\bibnamefont {Lu}}, \bibinfo {author} {\bibfnamefont
  {J.}~\bibnamefont {Yang}}, \bibinfo {author} {\bibfnamefont {X.}~\bibnamefont
  {Lu}}, \bibinfo {author} {\bibfnamefont {Q.}~\bibnamefont {Xiong}}, \ and\
  \bibinfo {author} {\bibfnamefont {H.}~\bibnamefont {Zhang}},\ }\href
  {\doibase 10.1021/nn4047474} {\bibfield  {journal} {\bibinfo  {journal} {ACS
  Nano}\ }\textbf {\bibinfo {volume} {7}},\ \bibinfo {pages} {10344} (\bibinfo
  {year} {2013})},\ \bibinfo {note} {pMID: 24131442},\ \Eprint
  {http://arxiv.org/abs/https://doi.org/10.1021/nn4047474}
  {https://doi.org/10.1021/nn4047474} \BibitemShut {NoStop}%
\bibitem [{\citenamefont {Zhao}\ \emph {et~al.}(2013)\citenamefont {Zhao},
  \citenamefont {Luo}, \citenamefont {Li}, \citenamefont {Zhang}, \citenamefont
  {Araujo}, \citenamefont {Gan}, \citenamefont {Wu}, \citenamefont {Zhang},
  \citenamefont {Quek}, \citenamefont {Dresselhaus},\ and\ \citenamefont
  {Xiong}}]{Zhao2013}%
  \BibitemOpen
  \bibfield  {author} {\bibinfo {author} {\bibfnamefont {Y.}~\bibnamefont
  {Zhao}}, \bibinfo {author} {\bibfnamefont {X.}~\bibnamefont {Luo}}, \bibinfo
  {author} {\bibfnamefont {H.}~\bibnamefont {Li}}, \bibinfo {author}
  {\bibfnamefont {J.}~\bibnamefont {Zhang}}, \bibinfo {author} {\bibfnamefont
  {P.~T.}\ \bibnamefont {Araujo}}, \bibinfo {author} {\bibfnamefont {C.~K.}\
  \bibnamefont {Gan}}, \bibinfo {author} {\bibfnamefont {J.}~\bibnamefont
  {Wu}}, \bibinfo {author} {\bibfnamefont {H.}~\bibnamefont {Zhang}}, \bibinfo
  {author} {\bibfnamefont {S.~Y.}\ \bibnamefont {Quek}}, \bibinfo {author}
  {\bibfnamefont {M.~S.}\ \bibnamefont {Dresselhaus}}, \ and\ \bibinfo {author}
  {\bibfnamefont {Q.}~\bibnamefont {Xiong}},\ }\href@noop {} {\bibfield
  {journal} {\bibinfo  {journal} {Nano Letters}\ }\textbf {\bibinfo {volume}
  {13}},\ \bibinfo {pages} {1007} (\bibinfo {year} {2013})}\BibitemShut
  {NoStop}%
\bibitem [{\citenamefont {Terrones}\ \emph {et~al.}(2014)\citenamefont
  {Terrones}, \citenamefont {Corro}, \citenamefont {Feng}, \citenamefont
  {Poumirol}, \citenamefont {Rhodes}, \citenamefont {Smirnov}, \citenamefont
  {Pradhan}, \citenamefont {Lin}, \citenamefont {Nguyen}, \citenamefont
  {El{\'i}as}, \citenamefont {Mallouk}, \citenamefont {Balicas}, \citenamefont
  {Pimenta},\ and\ \citenamefont {Terrones}}]{Terrones2014}%
  \BibitemOpen
  \bibfield  {author} {\bibinfo {author} {\bibfnamefont {H.}~\bibnamefont
  {Terrones}}, \bibinfo {author} {\bibfnamefont {E.~D.}\ \bibnamefont {Corro}},
  \bibinfo {author} {\bibfnamefont {S.}~\bibnamefont {Feng}}, \bibinfo {author}
  {\bibfnamefont {J.~M.}\ \bibnamefont {Poumirol}}, \bibinfo {author}
  {\bibfnamefont {D.}~\bibnamefont {Rhodes}}, \bibinfo {author} {\bibfnamefont
  {D.}~\bibnamefont {Smirnov}}, \bibinfo {author} {\bibfnamefont {N.~R.}\
  \bibnamefont {Pradhan}}, \bibinfo {author} {\bibfnamefont {Z.}~\bibnamefont
  {Lin}}, \bibinfo {author} {\bibfnamefont {M.~A.~T.}\ \bibnamefont {Nguyen}},
  \bibinfo {author} {\bibfnamefont {A.~L.}\ \bibnamefont {El{\'i}as}}, \bibinfo
  {author} {\bibfnamefont {T.~E.}\ \bibnamefont {Mallouk}}, \bibinfo {author}
  {\bibfnamefont {L.}~\bibnamefont {Balicas}}, \bibinfo {author} {\bibfnamefont
  {M.~A.}\ \bibnamefont {Pimenta}}, \ and\ \bibinfo {author} {\bibfnamefont
  {M.}~\bibnamefont {Terrones}},\ }\href@noop {} {\bibfield  {journal}
  {\bibinfo  {journal} {Scientific Reports}\ }\textbf {\bibinfo {volume} {4}},\
  \bibinfo {pages} {4215} (\bibinfo {year} {2014})}\BibitemShut {NoStop}%
\bibitem [{\citenamefont {Sheremetyeva}\ \emph {et~al.}(2018)\citenamefont
  {Sheremetyeva}, \citenamefont {Cherniak}, \citenamefont {Watson},\ and\
  \citenamefont {Meunier}}]{Sheremetyeva2018}%
  \BibitemOpen
  \bibfield  {author} {\bibinfo {author} {\bibfnamefont {N.}~\bibnamefont
  {Sheremetyeva}}, \bibinfo {author} {\bibfnamefont {D.~J.}\ \bibnamefont
  {Cherniak}}, \bibinfo {author} {\bibfnamefont {E.~B.}\ \bibnamefont
  {Watson}}, \ and\ \bibinfo {author} {\bibfnamefont {V.}~\bibnamefont
  {Meunier}},\ }\href {\doibase 10.1007/s00269-017-0906-1} {\bibfield
  {journal} {\bibinfo  {journal} {Physics and Chemistry of Minerals}\ }\textbf
  {\bibinfo {volume} {45}},\ \bibinfo {pages} {173} (\bibinfo {year}
  {2018})}\BibitemShut {NoStop}%
\bibitem [{\citenamefont {Kroumova}\ \emph {et~al.}(2003)\citenamefont
  {Kroumova}, \citenamefont {Aroyo}, \citenamefont {Perez-Mato}, \citenamefont
  {Kirov}, \citenamefont {Capillas}, \citenamefont {Ivantchev},\ and\
  \citenamefont {Wondratschek}}]{Kroumova2003}%
  \BibitemOpen
  \bibfield  {author} {\bibinfo {author} {\bibfnamefont {E.}~\bibnamefont
  {Kroumova}}, \bibinfo {author} {\bibfnamefont {M.~I.}\ \bibnamefont {Aroyo}},
  \bibinfo {author} {\bibfnamefont {J.~M.}\ \bibnamefont {Perez-Mato}},
  \bibinfo {author} {\bibfnamefont {A.}~\bibnamefont {Kirov}}, \bibinfo
  {author} {\bibfnamefont {C.}~\bibnamefont {Capillas}}, \bibinfo {author}
  {\bibfnamefont {S.}~\bibnamefont {Ivantchev}}, \ and\ \bibinfo {author}
  {\bibfnamefont {H.}~\bibnamefont {Wondratschek}},\ }\href {\doibase
  10.1080/0141159031000076110} {\bibfield  {journal} {\bibinfo  {journal}
  {Phase Transitions}\ }\textbf {\bibinfo {volume} {76}},\ \bibinfo {pages}
  {155} (\bibinfo {year} {2003})}\BibitemShut {NoStop}%
\bibitem [{\citenamefont {Chen}\ \emph {et~al.}(2015)\citenamefont {Chen},
  \citenamefont {Zheng}, \citenamefont {Fuhrer},\ and\ \citenamefont
  {Yan}}]{Chen2015}%
  \BibitemOpen
  \bibfield  {author} {\bibinfo {author} {\bibfnamefont {S.-Y.}\ \bibnamefont
  {Chen}}, \bibinfo {author} {\bibfnamefont {C.}~\bibnamefont {Zheng}},
  \bibinfo {author} {\bibfnamefont {M.~S.}\ \bibnamefont {Fuhrer}}, \ and\
  \bibinfo {author} {\bibfnamefont {J.}~\bibnamefont {Yan}},\ }\href@noop {}
  {\bibfield  {journal} {\bibinfo  {journal} {Nano Letters}\ }\textbf {\bibinfo
  {volume} {15}},\ \bibinfo {pages} {2526} (\bibinfo {year}
  {2015})}\BibitemShut {NoStop}%
\bibitem [{\citenamefont {Cao}\ \emph {et~al.}(2017)\citenamefont {Cao},
  \citenamefont {Sheremetyeva}, \citenamefont {Liang}, \citenamefont {Yuan},
  \citenamefont {Zhong}, \citenamefont {Meunier},\ and\ \citenamefont
  {Pan}}]{Cao2017}%
  \BibitemOpen
  \bibfield  {author} {\bibinfo {author} {\bibfnamefont {Y.}~\bibnamefont
  {Cao}}, \bibinfo {author} {\bibfnamefont {N.}~\bibnamefont {Sheremetyeva}},
  \bibinfo {author} {\bibfnamefont {L.}~\bibnamefont {Liang}}, \bibinfo
  {author} {\bibfnamefont {H.}~\bibnamefont {Yuan}}, \bibinfo {author}
  {\bibfnamefont {T.}~\bibnamefont {Zhong}}, \bibinfo {author} {\bibfnamefont
  {V.}~\bibnamefont {Meunier}}, \ and\ \bibinfo {author} {\bibfnamefont
  {M.}~\bibnamefont {Pan}},\ }\href@noop {} {\bibfield  {journal} {\bibinfo
  {journal} {2D Materials}\ }\textbf {\bibinfo {volume} {4}},\ \bibinfo {pages}
  {035024} (\bibinfo {year} {2017})}\BibitemShut {NoStop}%
\bibitem [{\citenamefont {Dong}\ \emph {et~al.}(2018)\citenamefont {Dong},
  \citenamefont {Xu}, \citenamefont {Zhang}, \citenamefont {Wu}, \citenamefont
  {Zhou}, \citenamefont {Liu}, \citenamefont {Dong}, \citenamefont {Fu},
  \citenamefont {Wu},\ and\ \citenamefont {Lei}}]{Dong2018}%
  \BibitemOpen
  \bibfield  {author} {\bibinfo {author} {\bibfnamefont {H.}~\bibnamefont
  {Dong}}, \bibinfo {author} {\bibfnamefont {Y.}~\bibnamefont {Xu}}, \bibinfo
  {author} {\bibfnamefont {C.}~\bibnamefont {Zhang}}, \bibinfo {author}
  {\bibfnamefont {Y.}~\bibnamefont {Wu}}, \bibinfo {author} {\bibfnamefont
  {M.}~\bibnamefont {Zhou}}, \bibinfo {author} {\bibfnamefont {L.}~\bibnamefont
  {Liu}}, \bibinfo {author} {\bibfnamefont {Y.}~\bibnamefont {Dong}}, \bibinfo
  {author} {\bibfnamefont {Q.}~\bibnamefont {Fu}}, \bibinfo {author}
  {\bibfnamefont {M.}~\bibnamefont {Wu}}, \ and\ \bibinfo {author}
  {\bibfnamefont {Y.}~\bibnamefont {Lei}},\ }\href@noop {} {\bibfield
  {journal} {\bibinfo  {journal} {Inorg. Chem. Front.}\ }\textbf {\bibinfo
  {volume} {5}},\ \bibinfo {pages} {3099} (\bibinfo {year} {2018})}\BibitemShut
  {NoStop}%
\bibitem [{\citenamefont {Xiao}\ \emph {et~al.}(2014)\citenamefont {Xiao},
  \citenamefont {Long}, \citenamefont {Li}, \citenamefont {Zhang},
  \citenamefont {Xu},\ and\ \citenamefont {Chan}}]{Xiao2014}%
  \BibitemOpen
  \bibfield  {author} {\bibinfo {author} {\bibfnamefont {J.}~\bibnamefont
  {Xiao}}, \bibinfo {author} {\bibfnamefont {M.}~\bibnamefont {Long}}, \bibinfo
  {author} {\bibfnamefont {X.}~\bibnamefont {Li}}, \bibinfo {author}
  {\bibfnamefont {Q.}~\bibnamefont {Zhang}}, \bibinfo {author} {\bibfnamefont
  {H.}~\bibnamefont {Xu}}, \ and\ \bibinfo {author} {\bibfnamefont {K.~S.}\
  \bibnamefont {Chan}},\ }\href@noop {} {\bibfield  {journal} {\bibinfo
  {journal} {Journal of Physics: Condensed Matter}\ }\textbf {\bibinfo {volume}
  {26}},\ \bibinfo {pages} {405302} (\bibinfo {year} {2014})}\BibitemShut
  {NoStop}%
\bibitem [{\citenamefont {Momma}\ and\ \citenamefont
  {Izumi}(2011)}]{Momma2011}%
  \BibitemOpen
  \bibfield  {author} {\bibinfo {author} {\bibfnamefont {K.}~\bibnamefont
  {Momma}}\ and\ \bibinfo {author} {\bibfnamefont {F.}~\bibnamefont {Izumi}},\
  }\href {\doibase 10.1107/S0021889811038970} {\bibfield  {journal} {\bibinfo
  {journal} {Journal of Applied Crystallography}\ }\textbf {\bibinfo {volume}
  {44}},\ \bibinfo {pages} {1272} (\bibinfo {year} {2011})}\BibitemShut
  {NoStop}%
\bibitem [{\citenamefont {Henkelman}\ \emph {et~al.}(2006)\citenamefont
  {Henkelman}, \citenamefont {Arnaldsson},\ and\ \citenamefont
  {{J\'onsson}}}]{Henkelman2006}%
  \BibitemOpen
  \bibfield  {author} {\bibinfo {author} {\bibfnamefont {G.}~\bibnamefont
  {Henkelman}}, \bibinfo {author} {\bibfnamefont {A.}~\bibnamefont
  {Arnaldsson}}, \ and\ \bibinfo {author} {\bibfnamefont {H.}~\bibnamefont
  {{J\'onsson}}},\ }\href {\doibase
  https://doi.org/10.1016/j.commatsci.2005.04.010} {\bibfield  {journal}
  {\bibinfo  {journal} {Computational Materials Science}\ }\textbf {\bibinfo
  {volume} {36}},\ \bibinfo {pages} {354 } (\bibinfo {year}
  {2006})}\BibitemShut {NoStop}%
\bibitem [{\citenamefont {Liang}\ \emph
  {et~al.}(2017{\natexlab{b}})\citenamefont {Liang}, \citenamefont {Puretzky},
  \citenamefont {Sumpter},\ and\ \citenamefont {Meunier}}]{Liang2017a}%
  \BibitemOpen
  \bibfield  {author} {\bibinfo {author} {\bibfnamefont {L.}~\bibnamefont
  {Liang}}, \bibinfo {author} {\bibfnamefont {A.~A.}\ \bibnamefont {Puretzky}},
  \bibinfo {author} {\bibfnamefont {B.~G.}\ \bibnamefont {Sumpter}}, \ and\
  \bibinfo {author} {\bibfnamefont {V.}~\bibnamefont {Meunier}},\ }\href@noop
  {} {\bibfield  {journal} {\bibinfo  {journal} {Nanoscale}\ }\textbf {\bibinfo
  {volume} {9}},\ \bibinfo {pages} {15340} (\bibinfo {year}
  {2017}{\natexlab{b}})}\BibitemShut {NoStop}%
\bibitem [{\citenamefont {Tristant}\ \emph {et~al.}(2018)\citenamefont
  {Tristant}, \citenamefont {Cupo},\ and\ \citenamefont
  {Meunier}}]{tristant2018finite}%
  \BibitemOpen
  \bibfield  {author} {\bibinfo {author} {\bibfnamefont {D.}~\bibnamefont
  {Tristant}}, \bibinfo {author} {\bibfnamefont {A.}~\bibnamefont {Cupo}}, \
  and\ \bibinfo {author} {\bibfnamefont {V.}~\bibnamefont {Meunier}},\
  }\href@noop {} {\bibfield  {journal} {\bibinfo  {journal} {2D Mater.}\
  }\textbf {\bibinfo {volume} {5}},\ \bibinfo {pages} {035044} (\bibinfo {year}
  {2018})}\BibitemShut {NoStop}%
\bibitem [{\citenamefont {Chakraborty}\ \emph {et~al.}(2012)\citenamefont
  {Chakraborty}, \citenamefont {Bera}, \citenamefont {Muthu}, \citenamefont
  {Bhowmick}, \citenamefont {Waghmare},\ and\ \citenamefont
  {Sood}}]{Chakraborty2012}%
  \BibitemOpen
  \bibfield  {author} {\bibinfo {author} {\bibfnamefont {B.}~\bibnamefont
  {Chakraborty}}, \bibinfo {author} {\bibfnamefont {A.}~\bibnamefont {Bera}},
  \bibinfo {author} {\bibfnamefont {D.~V.~S.}\ \bibnamefont {Muthu}}, \bibinfo
  {author} {\bibfnamefont {S.}~\bibnamefont {Bhowmick}}, \bibinfo {author}
  {\bibfnamefont {U.~V.}\ \bibnamefont {Waghmare}}, \ and\ \bibinfo {author}
  {\bibfnamefont {A.~K.}\ \bibnamefont {Sood}},\ }\href@noop {} {\bibfield
  {journal} {\bibinfo  {journal} {Phys. Rev. B}\ }\textbf {\bibinfo {volume}
  {85}},\ \bibinfo {pages} {161403} (\bibinfo {year} {2012})}\BibitemShut
  {NoStop}%
\bibitem [{\citenamefont {Zhang}\ \emph {et~al.}(2013)\citenamefont {Zhang},
  \citenamefont {Han}, \citenamefont {Wu}, \citenamefont {Milana},
  \citenamefont {Lu}, \citenamefont {Li}, \citenamefont {Ferrari},\ and\
  \citenamefont {Tan}}]{Zhang2013}%
  \BibitemOpen
  \bibfield  {author} {\bibinfo {author} {\bibfnamefont {X.}~\bibnamefont
  {Zhang}}, \bibinfo {author} {\bibfnamefont {W.~P.}\ \bibnamefont {Han}},
  \bibinfo {author} {\bibfnamefont {J.~B.}\ \bibnamefont {Wu}}, \bibinfo
  {author} {\bibfnamefont {S.}~\bibnamefont {Milana}}, \bibinfo {author}
  {\bibfnamefont {Y.}~\bibnamefont {Lu}}, \bibinfo {author} {\bibfnamefont
  {Q.~Q.}\ \bibnamefont {Li}}, \bibinfo {author} {\bibfnamefont {A.~C.}\
  \bibnamefont {Ferrari}}, \ and\ \bibinfo {author} {\bibfnamefont {P.~H.}\
  \bibnamefont {Tan}},\ }\href@noop {} {\bibfield  {journal} {\bibinfo
  {journal} {Phys. Rev. B}\ }\textbf {\bibinfo {volume} {87}},\ \bibinfo
  {pages} {115413} (\bibinfo {year} {2013})}\BibitemShut {NoStop}%
\bibitem [{\citenamefont {Xia}\ \emph {et~al.}(2017)\citenamefont {Xia},
  \citenamefont {Yan},\ and\ \citenamefont {Shen}}]{Xia2017}%
  \BibitemOpen
  \bibfield  {author} {\bibinfo {author} {\bibfnamefont {J.}~\bibnamefont
  {Xia}}, \bibinfo {author} {\bibfnamefont {J.}~\bibnamefont {Yan}}, \ and\
  \bibinfo {author} {\bibfnamefont {Z.~X.}\ \bibnamefont {Shen}},\ }\href
  {\doibase https://doi.org/10.1016/j.flatc.2017.06.007} {\bibfield  {journal}
  {\bibinfo  {journal} {FlatChem}\ }\textbf {\bibinfo {volume} {4}},\ \bibinfo
  {pages} {1 } (\bibinfo {year} {2017})}\BibitemShut {NoStop}%
\bibitem [{\citenamefont {Gong}\ \emph {et~al.}(2013)\citenamefont {Gong},
  \citenamefont {Huang}, \citenamefont {Miller}, \citenamefont {Cheng},
  \citenamefont {Hao}, \citenamefont {Cobden}, \citenamefont {Kim},
  \citenamefont {Ruoff}, \citenamefont {Wallace}, \citenamefont {Cho},
  \citenamefont {Xu},\ and\ \citenamefont {Chabal}}]{Gong2013}%
  \BibitemOpen
  \bibfield  {author} {\bibinfo {author} {\bibfnamefont {C.}~\bibnamefont
  {Gong}}, \bibinfo {author} {\bibfnamefont {C.}~\bibnamefont {Huang}},
  \bibinfo {author} {\bibfnamefont {J.}~\bibnamefont {Miller}}, \bibinfo
  {author} {\bibfnamefont {L.}~\bibnamefont {Cheng}}, \bibinfo {author}
  {\bibfnamefont {Y.}~\bibnamefont {Hao}}, \bibinfo {author} {\bibfnamefont
  {D.}~\bibnamefont {Cobden}}, \bibinfo {author} {\bibfnamefont
  {J.}~\bibnamefont {Kim}}, \bibinfo {author} {\bibfnamefont {R.~S.}\
  \bibnamefont {Ruoff}}, \bibinfo {author} {\bibfnamefont {R.~M.}\ \bibnamefont
  {Wallace}}, \bibinfo {author} {\bibfnamefont {K.}~\bibnamefont {Cho}},
  \bibinfo {author} {\bibfnamefont {X.}~\bibnamefont {Xu}}, \ and\ \bibinfo
  {author} {\bibfnamefont {Y.~J.}\ \bibnamefont {Chabal}},\ }\href@noop {}
  {\bibfield  {journal} {\bibinfo  {journal} {ACS Nano}\ }\textbf {\bibinfo
  {volume} {7}},\ \bibinfo {pages} {11350} (\bibinfo {year}
  {2013})}\BibitemShut {NoStop}%
\bibitem [{\citenamefont {Freedy}\ \emph {et~al.}(2019)\citenamefont {Freedy},
  \citenamefont {Zhang}, \citenamefont {Litwin}, \citenamefont {Bendersky},
  \citenamefont {Davydov},\ and\ \citenamefont {McDonnell}}]{Freedy2019}%
  \BibitemOpen
  \bibfield  {author} {\bibinfo {author} {\bibfnamefont {K.~M.}\ \bibnamefont
  {Freedy}}, \bibinfo {author} {\bibfnamefont {H.}~\bibnamefont {Zhang}},
  \bibinfo {author} {\bibfnamefont {P.~M.}\ \bibnamefont {Litwin}}, \bibinfo
  {author} {\bibfnamefont {L.~A.}\ \bibnamefont {Bendersky}}, \bibinfo {author}
  {\bibfnamefont {A.~V.}\ \bibnamefont {Davydov}}, \ and\ \bibinfo {author}
  {\bibfnamefont {S.}~\bibnamefont {McDonnell}},\ }\href {\doibase
  10.1021/acsami.9b08829} {\bibfield  {journal} {\bibinfo  {journal} {ACS
  Applied Materials and Interfaces}\ }\textbf {\bibinfo {volume} {11}},\
  \bibinfo {pages} {35389} (\bibinfo {year} {2019})}\BibitemShut {NoStop}%
\bibitem [{\citenamefont {Walter}\ \emph {et~al.}(2020)\citenamefont {Walter},
  \citenamefont {Cooley}, \citenamefont {Domask},\ and\ \citenamefont
  {Mohney}}]{Walter2020}%
  \BibitemOpen
  \bibfield  {author} {\bibinfo {author} {\bibfnamefont {T.~N.}\ \bibnamefont
  {Walter}}, \bibinfo {author} {\bibfnamefont {K.~A.}\ \bibnamefont {Cooley}},
  \bibinfo {author} {\bibfnamefont {A.~C.}\ \bibnamefont {Domask}}, \ and\
  \bibinfo {author} {\bibfnamefont {S.~E.}\ \bibnamefont {Mohney}},\ }\href
  {\doibase 10.1016/j.mssp.2019.104850} {\bibfield  {journal} {\bibinfo
  {journal} {Materials Science in Semiconductor Processing}\ }\textbf {\bibinfo
  {volume} {107}},\ \bibinfo {pages} {104850} (\bibinfo {year}
  {2020})}\BibitemShut {NoStop}%
\end{thebibliography}%
\end{document}